\documentstyle[12pt]{article}
\makeatletter
\newdimen\normalarrayskip              
\newdimen\minarrayskip                 
\normalarrayskip\baselineskip
\minarrayskip\jot
\newif\ifold             \oldtrue            \def\new{\oldfalse}
\def\arraymode{\ifold\relax\else\displaystyle\fi} 
\def\eqnumphantom{\phantom{(\theequation)}}     
\def\@arrayskip{\ifold\baselineskip\z@\lineskip\z@
     \else
     \baselineskip\minarrayskip\lineskip2\minarrayskip\fi}
\def\@arrayclassz{\ifcase \@lastchclass \@acolampacol \or
\@ampacol \or \or \or \@addamp \or
   \@acolampacol \or \@firstampfalse \@acol \fi
\edef\@preamble{\@preamble
  \ifcase \@chnum
     \hfil$\relax\arraymode\@sharp$\hfil
     \or $\relax\arraymode\@sharp$\hfil
     \or \hfil$\relax\arraymode\@sharp$\fi}}
\def\@array[#1]#2{\setbox\@arstrutbox=\hbox{\vrule
     height\arraystretch \ht\strutbox
     depth\arraystretch \dp\strutbox
     width\z@}\@mkpream{#2}\edef\@preamble{\halign \noexpand\@halignto
\bgroup \tabskip\z@ \@arstrut \@preamble \tabskip\z@ \cr}%
\let\@startpbox\@@startpbox \let\@endpbox\@@endpbox
  \if #1t\vtop \else \if#1b\vbox \else \vcenter \fi\fi
  \bgroup \let\par\relax
  \let\@sharp##\let\protect\relax
  \@arrayskip\@preamble}
\def\eqnarray{\stepcounter{equation}%
              \let\@currentlabel=\theequation
              \global\@eqnswtrue
              \global\@eqcnt\z@
              \tabskip\@centering
              \let\\=\@eqncr
              $$%
 \halign to \displaywidth\bgroup
    \eqnumphantom\@eqnsel\hskip\@centering
    $\displaystyle \tabskip\z@ {##}$%
    &\global\@eqcnt\@ne \hskip 2\arraycolsep
         \hfil$\arraymode{##}$\hfil
    &\global\@eqcnt\tw@ \hskip 2\arraycolsep
         $\displaystyle\tabskip\z@{##}$\hfil
         \tabskip\@centering
    &{##}\tabskip\z@\cr}
\makeatother

\def\beq{\begin{equation}}
\def\eeq{\end{equation}}
\def\bea{\begin{eqnarray}}
\def\eea{\end{eqnarray}}
\def\nn{\nonumber}
\def\stackreb#1#2{\mathrel{\mathop{#2}\limits_{#1}}}
\def\theequation{\thesection.\arabic{equation}}  

\begin{document}

\begin{titlepage}
\begin{center}
{{\it P.N.Lebedev Institute preprint} \hfill FIAN/TD-10/91\\
{\it I.E.Tamm Theory Department} \hfill ITEP-M-9/91
\begin{flushright}{October 1991}\end{flushright}
\vspace{0.1in}{\Large\bf Towards unified theory of $2d$ gravity}\\[.4in]
{\large  S.Kharchev, A.Marshakov, A.Mironov}\\
\bigskip {\it Department of Theoretical Physics \\  P.N.Lebedev Physical
Institute \\ Leninsky prospect, 53, Moscow, 117 924},
\footnote{E-mail address: theordep@sci.fian.msk.su}\\ \smallskip
\bigskip {\large A.Morozov}\\
 \bigskip {\it Institute of Theoretical and Experimental
Physics,  \\
 Bol.Cheremushkinskaya st., 25, Moscow, 117 259},
 \footnote{E-mail address: morozov@itep.msk.su}\\ \smallskip
\bigskip {\large A.Zabrodin}\\
 \bigskip {\it Institute of Chemical Physics\\ Kosygina st.,4, 117334, Moscow}}
\end{center}
\bigskip
\bigskip

\newpage
\setcounter{page}2
\centerline{\bf ABSTRACT}
\begin{quotation}

We introduce a new 1-matrix model with arbitrary potential and the
matrix-valued background field. Its partition function is a $\tau$-function of
KP-hierarchy, subjected to a kind of
${\cal L}_{-1}$-constraint. Moreover, partition function behaves smoothly in
the limit of infinitely large matrices. If the potential is equal to
$X^{K+1}$, this partition function becomes a $\tau$-function of $K$-reduced
KP-hierarchy, obeying a set of ${\cal W} _K$-algebra constraints identical to
those conjectured in \cite{FKN91} for double-scaling continuum limit of
$(K-1)$-matrix model. In the case of $K=2$ the statement reduces to the early
established \cite{MMM91b} relation between Kontsevich model and the ordinary
$2d$ quantum gravity . Kontsevich model with generic potential may be
considered as interpolation between all the models of $2d$ quantum gravity
with $c<1$ preserving the property of integrability and
the analogue of string equation.
\end{quotation}
\end{titlepage}

{\bf CONTENTS:}\\
1. Motivation and results

1.1. Introduction

1.2. Generalized Kontsevich model

1.3. Properties of $Z^{\{2\}}$

1.4. The case of arbitrary $K$

1.5. Comments\\
2. Partition function of GKM as KP $\tau $-function

2.1. Short summary

2.2. From GKM to determinant formula

2.3. KP $\tau $-function in Miwa parameterization

2.4. Hirota equations for $\tau $-function in Miwa coordinates

2.5. Grassmannian description of $\tau $-function

2.6. GKM and reductions of KP

2.7 On $T_{nK}$-dependence of $Z^{\{K\}}$\\
3. Derivation of the ${\cal L}_{-1}$-constraint

3.1. Motivations

3.2. Direct evaluation of ${\cal L}_{-1}Z$

3.3 Universal string equation

3.4 Discussion\\
4. From Ward identities to ${\cal W}$-constraints

4.1. General discussion

4.2. Virasoro constraints in Kontsevich model $(K=2)$

4.3. Example of $K=3$\\
5. Conclusion\\
6. Acknowledgments\\
7. References

\newpage
\setcounter{footnote}0
\section{Motivation and results}
\subsection{Introduction}

Matrix models play an increasingly important role in the theory of quantum
gravity, being a reasonably simple realization of Regge calculus, adequate at
least in essentially to topological theories. Their conventional application to
string theory is based on identification of the critical points of peculiar
``double-scaling" limit \cite{Kaz89b,BK90,DS90,GM90a} of Hermitean multimatrix
models \cite{CMM81}
with the $c<1$ minimal conformal field theories \cite{BPZ84} coupled to $2d$
gravity.
Such models are often supposed \cite{Dou90} to exhaust the entire set of
consistent
bosonic 2-dimensional string models, while bosonic strings beyond 2 space-time
dimensions are presumably unstable \cite{Sei90}. Technical methods, based on
the
application of matrix models, appeared rather successful not only in the
framework of perturbation theory, but allow one to estimate the entire sums of
perturbation series, thus rising the study of string models to a qualitatively
new level.

On the other hand, matrix models appeared to possess unexpectedly deep internal
mathematical structure \cite{Dou90}. In fact, if matrix models are used to
describe
interpolation between different critical points $(i.e$. particular $c<1$ string
models), the interpolating flows are mutually commuting, $i.e$. the entire
pattern of flows constitutes an integrable hierarchy (some reduction of
Kadomtsev-Petviashvili (KP) system). This integrable structure indicates that
the somewhat artificial description in terms of matrix integrals should possess
a more invariant algebro-geometrical description, which could be used for
formulation of the yet unknown dynamical principle of string theory, unifying
all available string models into distinguished and (hopefully) physically
important entity.

\bigskip

The main obstacles on the way from matrix models to such general principle are:

(i) the sophisticated intermediate step: the double-scaling limit,

(ii) the somewhat obscure origin of integrability: it is not {\it a priori}
obvious, why matrix integrals with generic ``potentials" are $\tau $-functions
of (reduced) KP-hierarchies,

(iii) the non-universal description of different $c<1$ string models: some
of them are nicely unified as different critical points of particular
multimatrix model, but there is no nice interpolation between critical points
of models with the different number of matrices. The naive way out is to use
$\infty $-matrix model, but it is no longer described in terms of
{\it finite}-dimensional matrix integrals, and, more important, reductions
to the
case of particular multimatrix model is too much singular.

\bigskip

In this paper we propose a kind of resolution of all these three problems.
Namely, we argue that there is a new {\it one}-matrix model \cite{KMMMZ91a}
(which we
call {\it Generalized Kontsevich Model} (GKM)), which

(i) for a specially adjusted potential describes properly the double scaling
limit of any multimatrix model,

(ii) has the partition function, which is a KP $\tau $-function, properly
reducible at the points, associated with multimatrix models, and satisfies an
additional equation, which reduces to conventional ${\cal L}_{-1}$-constraints
for multimatrix models,

(iii) allows continuous deformation of potential, changing it from the form,
associated with a given multimatrix model to those corresponding to the others.

\bigskip

This makes the study of GKM a natural step in the development of string theory.

In the remaining part of this section we describe GKM and its relation to
multimatrix models more explicitly. The proofs are presented in sections 2-4.
For a more condensed presentation of our results see \cite{KMMMZ91a}.

\subsection{Generalized Kontsevich model}

This paper is devoted to the study of 1-matrix model in external matrix field
$\Lambda $, essentially defined by the integral over $N\times N$
matrix $X:$

\beq
{\cal F}^{\{{{\cal V}}\}}_N[\Lambda ] \equiv  \int   dX\ e^{-Tr{{\cal V}}(X) +
Tr\Lambda X}
\eeq
with arbitrary ``potential" ${{\cal V}}(X)$ and $dX=\prod _{i,j=1}^NdX_{ij}$.
Such integrals arise in
consideration of various problems \cite{GN91,MS91,MMM91b,MMM91c}, but
{\it this time}
our purpose is to use (1.1) in order to define a {\it universal matrix model},
which could describe interpolation between {\it all} the multimatrix models and
their critical points, as explained in sect.1.1 above. Namely, we present
strong arguments that partition function of such universal model may be
identified with
$Z^{\{{{\cal V}}\}}_\infty [M]$, where

\beq
Z^{\{{{\cal V}}\}}_N[M] \equiv  {\int e^{-U(M,Y)}{dY}\over
\int e^{-U_2(M,Y)}dY}
\eeq
with
\beq
U(M,Y) = Tr[{{\cal V}}(M+Y) - {{\cal V}}(M) - {{\cal V}}'(M)Y]
\eeq
and

\beq
U_2(M,Y) =  \lim _{\epsilon \rightarrow 0}
{1\over \epsilon ^2} U(M,\epsilon Y)\hbox{,}
\eeq
$(i.e$. $U_2(M,Y)$ is the $Y^2$-contribution to $U(M,Y)).$

After the shift of integration variable

\beq
X = M+Y
\eeq
the numerator in (1.2) may be rewritten in terms of (1.1):

\beq
\int   e^{-U(M,Y)}dY = e^{Tr[{{\cal V}}(M)-M{{\cal V}} '(M)]}
{\cal F}^{\{{{\cal V}}\}}_N[{{\cal V}} '(M)]\hbox{.}
\eeq
The denominator in (1.2) is a simple Gaussian integral --- a peculiar matrix
generalization of  $[{{\cal V}} ''(\mu )]^{-1/2}.$

We shall call $Z^{\{{{\cal V}}\}}_N[M]$, defined by eq.(1.2), the partition
function of {\it generalized Kontsevich model} (GKM). The reason for this is
that for the special choice of potential\footnote{
We do not specify integration contour in matrix integrals (1.1), (1.2), since
all what we are going to discuss does not depend on it. To make contact with
the
literature, note that for potential ${{\cal V}}_K=
{X^{K+1}\over K+1}$ integrals should be over Hermitean or antiHermitean
matrices
if $K+1$ is even or odd respectively.}:

\beq
{{\cal V}}(X) = {{\cal V}}_2(X) = X^3/3,
\eeq
eq.(1.2) becomes the partition function of original Kontsevich model
\cite{Kon91}:

\bea
Z^{\{2\}}_N[M] = {\int dY\ e^{-1/3\ TrY^3 - TrMY^2}\over
\int dY\ e^{-TrMY^2}} = \nn \\
= e^{-(2/3)TrM^3} {\int dX\ e^{-1/3\ TrX^3 + TrM^2X}\over
\int dX\ e^{-TrMX^2}}\ .
\eea
Eq.(1.8) has a natural generalization, which is also a particular case of
(1.2): if

\beq
{{\cal V}}(X) = {{\cal V}}_K(X) = {X^{K+1}\over K+1}
\eeq
eq.(1.2) becomes

\beq
Z^{\{K\}}_N[M] \equiv  {e^{- {K\over K+1}TrM^{K+1}}{\int dX\ e^{-Tr{X^{K+1}
\over K+1} + TrM^KX}}\over \int dX\ e^{-{1\over K}Tr[\sum _{a+b=K-1}
M^aXM^bX]}}
\hbox{ . }
\eeq
We shall argue that $Z^{\{K\}}_\infty [M]$, defined by (1.10), is identical to
the square root of partition function of $(K-1)$-matrix model in the
double-scaling limit with

\beq
T_n = {1\over n} Tr\ M^{-n}\hbox{, }   n\neq 0\ \ mod\ K\ ,
\eeq
playing the role of generalized Kazakov's time-variables \cite{Kaz89b,MMMM91}:

\beq
Z^{\{K\}}_\infty [M] = \sqrt{\Gamma ^{\{K-1\}}_{ds}(T_n)},
\eeq
where \cite{CMM81}

\beq
\Gamma ^{\{K\}}_{ds} =\lim _{double\hbox{  }scaling} \prod _{i=1}^K
 \int   dM_i e^{\sum \ t_nTrM^n_i + TrM_iM_{i+1}}
\eeq
(the term $M_KM_{K+1}$ in the exponent should be omitted; the change from
discrete time-variables $\{t_n\}$ to continuum $\{T_n\}$ is implicit in
performing the double-scaling limit, see \cite{MMMM91,GMMMMO91}).

Since potential  ${{\cal V}}(X)$  can be continuously deformed from one $K$ in
(1.9) to another, the GKM (1.2) as a corollary of (1.12) may be considered as
continuous interpolation between all the models of $2d$ gravity with $c<1$,
described by particular multicritical points of particular multimatrix models
(let us remind, that for $K$ fixed the $p$-th multicritical point is defined by
the condition \cite{FKN91} that

\beq
\hbox{all }T_n=0\hbox{, except for }T_1\hbox{ and }T_{K+p},
\eeq
and according to (1.11) this is a constraint on the form of the matrix $M).$

This is the sense in which the GKM (1.2) may be considered as the
universal
partition function of bosonic string models. We shall not discuss the
implications of this fact here, instead we shall concentrate on the arguments
in favour of this conclusion, which are entirely based on the identity (1.12)
(and also motivated by the nice features of interpolating function
$Z^{\{{{\cal V}}\}}_\infty [M]$ itself --- it is usually a KP $\tau $-function,
subjected to ${\cal L}_{-1}$-constraint, see below). Our strategy is to try to
understand as much as possible about $Z^{\{K\}}_\infty [M]$, which stands at
the $l.h.s$. of (1.12) and compare these results with what is conjectured in
\cite{FKN91} about $\sqrt{\Gamma ^{\{K-1\}}_{ds}}$ at the $r.h.s.$

\subsection{Properties of $Z^{\{2\}}$}

In order to motivate our considerations let us remind first, what is known
about $Z^{\{K\}}[M]$ in the particular case of $K=2.$

Expression (1.2) has been derived in \cite{Kon91} as a representation of the
generating
functional of intersection numbers of the stable cohomology classes on the
universal module space, $i.e$. it is defined to be a partition function of
Witten's $2d$ topological gravity \cite{Wit90}. In \cite{MMM91b} (see also
\cite{MS91,Wit91} for
alternative derivations) it was shown that as $N \rightarrow  \infty \ $
$Z^{\{2\}}_\infty $ considered as a function of time variables (1.11),
satisfies
the set of Virasoro constraints

\beq
{\cal L}^{\{K=2\}}_nZ^{\{2\}}_\infty  = 0, \ \ \ n \geq -1\ ,
\eeq

\bea
{\cal L}^{\{2\}}_n= {1\over 2} \sum _{k\ odd} kT_k\partial /\partial T_{k+2n}+
{1\over 4}
\sum _{{a+b=2n}\atop {a,b\ odd\ and>0}}\partial ^2/
\partial T_a\partial T_b+\nn \\
+ {1\over 4}
\sum _{{a+b=-2n}\atop {a,b\ odd\ and>0}}aT_abT_b+ {1\over 16}\delta _{n,0} -
\partial /\partial T_{3+2n}.
\eea
(Notations are slightly changed as compared to \cite{MMM91b}. This derivation
is
partly reproduced in sect.4.2 below.). Sometimes it may be convenient to
consider the appearance of the last item at the $r.h.s$. of (1.16) as result of
additional shift of time-variables,

\beq
T_n \rightarrow  \hat T^{\{K\}}_n \equiv  T_n - {K\over n}\delta _{n,K+1},
\eeq
however, these $\hat T$-times are defined in a $K$-dependent way and do not
seem to have too much sense. Constraints (1.15) are exactly the equations
\cite{FKN91,DVV91a,MMMM91}, imposed on  $\sqrt{\Gamma ^{\{1\}}_{ds}}$.
Moreover, it is
very plausible that they possess a unique solution, so that (1.15) in fact
implies that

\beq
Z^{\{2\}}_\infty  = \sqrt{\Gamma ^{\{1\}}_{ds}}.
\eeq
Alternative solution to the constraints (1.15) is given \cite{FKN91,DVV91a}
by a
Galilean-invariant KdV $\tau $-function \cite{Orl88}, and, relying upon the
same
belief that the solution is unique, one concludes that  $Z^{\{2\}}_\infty $ is
a $\tau $-function of KdV-hierarchy, subjected to additional
${\cal L}^{\{2\}}_{-1}$-constraint:

\beq
Z^{\{2\}}_\infty  = \tau _{KdV} = \tau ^{\{2\}},
\eeq

\beq
{\cal L}^{\{2\}}_{-1}\tau ^{\{2\}} = 0.
\eeq
Notation $\tau ^{\{2\}}$ reflects the fact that any $\tau _{KdV}$ may be
considered as a KP $\tau $-function, evaluated at specific points of
Grassmannian (see sect.2.5 below for details), corresponding to the 2-reduction
of KP hierarchy. Generically, the $\tau $-function of $K$-reduced hierarchy,
$\tau ^{\{K\}}(T_n)$, is the KP $\tau $-function at peculiar points of
Grassmannian and possesses the following property:

\bea
\tau ^{\{K\}}(T_1...T_{K-1},T_K,T_{K+1}...T_{2K-1},T_{2K},T_{2K+1},...) =\nn \\
= e^{(\sum _na_{nK}T_{nK})}
\tau ^{\{K\}}(T_1...T_{K-1},0,T_{K+1}...T_{2K-1},0,T_{2K+1},...).
\eea
This means that the entire dependence of generic $\tau ^{\{K\}}$ of all
variables $T_{nK}$ ($i.e${\it . $T_n$} with $n=0$ {\it mod $K)$} is exhausted
by
the simple exponent

\beq
\exp (\sum  _na_{nK}T_{nK})
\eeq
with time-independent parameters $a_{nK}$. Moreover, this factor is
actually absent in
the case of GKM partition function: $a_{nK}[Z^{\{K\}}] =0.$

The significance of the just reported results concerning $K=2$ is two-fold.
First, they establish the relation (1.7) between two different models of $2d$
gravity: Witten's topological gravity \cite{Wit91}, represented by
$Z^{\{2\}}_\infty $
\cite{Kon91}, and ordinary quantum $2d$ gravity \cite{BK90,DS90,GM90a},
represented by
$\Gamma ^{\{1\}}_{ds}$. This was the main emphasize of ref.\cite{MMM91b}. The
second implication, and it is the one of interest for us in this paper, is the
possibility to describe the sophisticated double-scaling limit of matrix model
in terms of a completely different matrix model (1.2). It is important that
$Z^{\{K\}}_N$ is in fact a ``smooth" function of $N$ as $N \rightarrow
\infty $, and all the properties of $Z^{\{K\}}_\infty $ are just the same as at
finite $N$'s, --- in variance with the double-scaling limit for ordinary matrix
models, the limit $N \rightarrow  \infty $ for GKM is non-singular. Also, to
avoid misunderstanding, we should emphasize that what we mean under
double-scaling limit here is not the limit, describing a particular fixed
point, but the entire pattern of all the flows between various fixed points of
a given $(K-1)$-matrix model, so that partition function $\Gamma ^{\{K\}}_{ds}$
is indeed a function of all time-variables (1.3).

\subsection{The case of arbitrary $K$}

In order to generalize the identity (1.15) to the case of arbitrary $K$,
$i.e$. to prove (1.12), one can try to prove either an analogue of (1.15) or
that of (1.19), (1.20) and compare the result with the what is expected for
$\sqrt{\Gamma ^{\{K-1\}}_{ds}}$. This will be our strategy below. Namely, in
sect.2 a simple and promising formalism is developed, based on interpretation
of
(1.11) as Miwa transformation, and we use it to prove that {\it any
$Z^{\{{{\cal V}}\}}_\infty [M]$}, defined by (1.2) and (1.11) $(i.e$. with
arbitrary ${{\cal V}}(X))$ is in fact a KP $\tau $-function --- in particular,
satisfies bilinear Hirota difference equations. Moreover, in the Grassmannian
parameterization of KP $\tau $-functions the point of Grassmannian is specified
by the choice of potential ${{\cal V}}(X)$, so that multimatrix models with
different $K$ are (due to (1.12)) associated with different points of
Grassmannian. (In terms of the universal module space these points are in fact
associated with certain infinite-genus hyperellyptic surface for $K=2$, and
with certain infinite genus abelian coverings of degrees $K$ in the general
case. In this sense interpolation with the changing ${{\cal V}}(X)$ is some
flow
at infinity of the universal module space.)

What is specific for a given $K$ is that whenever ${{\cal V}}(X)$ is a
homogeneous polynomial of degree $K+1$, $i.e$. if it is of the form (1.9), the
KP $\tau $-function becomes independent of all $T_{Kn}$, $i.e$. acquires the
property (1.21) and may be considered as $\tau $-function $\tau ^{\{K\}}$ of
$K$-reduced hierarchy. This will be also explained in more details in sect.2
below.

Though any $Z^{\{{{\cal V}}\}}_\infty [M]$ defined by (1.2) is a KP
$\tau $-function, the inverse statement is not correct.
$Z^{\{{{\cal V}}\}}_\infty [M]$ is subjected to the infinite set of
constraints,
essentially implied by the Ward identities for the integral (1.1)
\cite{GN91,MMM91b}:

\beq
\left\lbrace Tr\epsilon (\Lambda )\left [{{\cal V}}'({\partial \over
\partial \Lambda _{tr}}) - \Lambda \right ]\right\rbrace
{\cal F}^{\{{{\cal V}}\}}_N =  0
\eeq
with any $\epsilon (\Lambda )$. These identities result from invariance of
(1.1)
under any shift of integration variable   $X \rightarrow  X +
\epsilon (\Lambda )$. Eq.(1.23) seems to provide a complete set of differential
equations, unambiguously defining ${\cal F}^{\{{{\cal V}}\}}_N$ (up to a
$\Lambda $ and $N$-independent constant). This statement is the actual basis
for all of the arguments, which make use of the uniqueness of solutions to any
constraints.

As shown in \cite{MMM91b} in the particular case of $K=2$ the entire set of
such
identities with all possible $\epsilon (\Lambda )$ is equivalent to (1.15) and
thus implies (1.19) and (1.20). In fact in sect.2 we follow a somewhat
different
logic and give an alternative {\it direct} derivation of relations

\beq
Z_{\infty}^{\{{{\cal V}}\}}=\tau \hbox{ , } \ \ \
Z_{\infty}^{\{K\}}=\tau ^{\{K\}}
\eeq
for any ${{\cal V}}(X)$ and $K$, so it remains only to prove the analogue of
(1.20). Such direct proof is presented in sect.3, moreover it is valid not only
for potentials of the form (1.9), but for {\it any ${{\cal V}}(X):$}

\bea
{\cal L}^{\{{{\cal V}}\}}_{-1}Z^{\{{{\cal V}}\}} = \left\{\sum _{n\geq 1}Tr
[{1\over {{\cal V}}''(M)M^{n+1}}] {\partial \over \partial T_n} \right.
+\nn \\
+ \left. {1\over 2}
\sum _{i,j}{1\over {{\cal V}}''(\mu _i){{\cal V}}''(\mu _j)}{{{\cal V}}''
(\mu _i)-
{{\cal V}}''(\mu _j)\over \mu _i - \mu  _j} -
{\partial \over \partial T_1} \right\}Z^{\{{{\cal V}}\}} = 0\ ,
\eea
where $\mu _i$ are eigenvalues of the matrix $M$.

For potentials (1.9)  ${\cal L}^{\{{{\cal V}}\}}_{-1}$ acquires conventional
form
of ${\cal L}^{\{K\}}_{-1}$ (see eq.(1.28) below) and may be easily supplemented
by other generators of Virasoro algebra.

Despite of eqs.(1.24) and (1.25) provide a complete description of partition
function of GKM, the study of direct consequences of Ward identities (1.23)
also seems interesting. In particular, it should lead to a direct derivation of
the analogue of Virasoro constraints (1.15). It also deserves noting that the
meaning of constraints on $Z^{\{{{\cal V}}\}}_\infty [M]$ is that
${{\cal V}}(X)$
parametrizes only some restricted subset of the universal module space. Only
this subset is associated with matrix models and interpolations between them.
This is also clear from the fact that  ${{\cal V}}(X) = \sum \ {\cal V}_nX^n$
is
naturally parameterized by an infinite vector $\{{\cal V}_n\}$ rather than
by a matrix
$A_{mn}$, what would be natural for the entire Grassmannian. Remarkably enough
the ``dimension" of this subset is just the same as that of the space of
time-variables (the space of potentials is parameterized by a vector
$\{{\cal V}_n\}$,
while that of time-variables --- by $\{T_n\}$ or by eigenvalue vector
$\{\mu _n\}$ of the matrix $M)$. This clearly suggests that a formalism should
exist for so restricted set of $\tau $-functions, which would treat potentials
and times in a more symmetrical fashion (see also discussion in
sect.3.4)\footnote{Let us also remark that such
symmetry exposes itself in the fact that the degree of potential $K$ and the
order of the multicriticity $p+K$ (fixed by the choice of times) give
the theory of minimal
matter with $c=1-{6(p-K)^2 \over pK}$ coupled to $2d$ gravity, and, thus,
$p$ and $K$ should  enter the
theory on equal footing.}.

Implications of Ward identities (1.23) for $Z^{\{K\}}_\infty [M]$ with $K>2$
are discussed to some extent in sect.4. This is a straightforward calculation,
though much more tedious than in the simplest case of $K=2$. We argue that the
proper generalization of (1.15) for the case of $K=3$ is given by the following
set of constraints:

\beq
\new
\begin{array}{c}
{{\cal W}}^{(3)}_{3n}Z^{\{3\}}_\infty  = 0\hbox{, }     n\geq -2\hbox{;}\\
\left\lbrace \sum _{k\geq 1}(3k-1)\hat T_{3k-1}{\cal W} ^{(2)}_{3k+3n}
+\sum _{a+b=3n} {\partial \over \partial T_{3a+2}} {\cal W}
^{(2)}_{3b-3}\right\rbrace  Z^{\{3\}}_\infty  = 0\hbox{, }  a,b\geq 0\hbox{, }
n\geq -2;\\
\left\lbrace \sum _{k\geq 1+\delta _{n+3,0}}(3k-2)\hat T_{3k-2}
{\cal W} ^{(2)}_{3k+3n} +\sum _{a+b=3n} {\partial \over \partial T_{3a+1}}
{\cal W} ^{(2)}_{3b-3}\right\rbrace  Z^{\{3\}}_\infty  = 0\hbox{, }
a,b\geq 0\hbox{, }  n\geq -3.
\end{array}
\eeq
Here ${\cal W} ^{(j)}_{Kn}$ stands for the $Kn$-th harmonics of the $j$-th
generator of Zamolodchikov's ${\cal W}_K$-algebra, and the
proper notation would be
${\cal W}^{(j)\{K\}}_n$ or ${\cal W}^{(j)\{K\}}_{Kn}$. The way to express these
quantities through free fields is described in \cite{FKN91}; for example,

\beq
\new
\begin{array}{l}
{\cal W}^{(3)}_{3n}  =
3\sum _{k,l\geq 1}k\hat T_kl\hat T_l{\partial \over \partial T_{k+l+3n}} +
3\sum _{k\geq 1}k\hat T_{k}\sum _{a+b=k+3n}
{\partial ^2\over \partial T_a\partial T_b}
+\sum _{a+b+c=3n} {\partial ^3\over \partial T_a\partial T_b\partial T_c}+\nn
\\
+ \sum _{a+b+c=-3n}a\hat T_ab\hat T_bc\hat T_c\hbox{; }    a,b,c >
0\hbox{, }  a,b,c,k,l \neq  0\ \ mod\ 3
\end{array}
\eeq
and

\bea
{\cal W}^{(2)}_{Kn} \equiv  {\cal L}^{\{K\}}_n = {1\over K} \sum _k
k\hat T_k\partial /\partial T_{k+Kn} + {1\over 2K }
\sum _{{a+b=Kn}\atop {a,b>0}}\partial ^2/\partial T_a\partial T_b +\nn \\
+ {1\over 2K }
\sum _{{a+b=-Kn}\atop {a,b>0}}a\hat T_ab\hat T_b +
{(K-1)(K+1)\over 24K}\delta _{n,0}.
\eea
(Note that in order to make formulas a bit more compact, we expressed all these
operators through  $\hat T^{\{3\}}_n$ rather than $T_n$-variables. These are
defined in (1.17), $\partial /\partial \hat T = \partial /\partial T.)$ The
analogue of (1.26) for $K>3$ is a more or less obvious generalization of
(1.26), involving all the generators  ${\cal W}^{(K)}_{Kn}$,
${\cal W}^{(K-1)}_{Kn}$, ..., ${\cal W}^{(2)}_{Kn}$ of ${\cal W}_K$-algebra
with the
``coefficients" made from $\hat T$'s and $(\partial /\partial T)$'s. The form
of these operators is a bit reminiscent of the $\tilde {\cal W}$-operators,
introduced in \cite{MMM91c,Fut} as ingredients of Ward identities in
{\it discrete} multimatrix models.

Eqs.(1.26) (and their analogues for $K>3)$ are obviously resolved by any
solution of the simpler set of equations,

\beq
{\cal W}^{(k)}_{Kn} Z^{\{K\}} = 0\hbox{, }   k = 2,3,...,K\hbox{; }  n\geq  1-k
\eeq
(eq.(1.20) is the particular case: $k=2$, $n = -1)$. On the other hand (1.26)
are equivalent to (1.23) for the particular form of potential (1.9), and thus
are supposed to possess a single solution. Therefore from this point of view it
also looks plausible that (1.26) (and its counterpart for any $K)$ is simply
equivalent to (1.29). Unfortunately we do not provide an explicit proof of this
equivalence (though we could rely, of course, upon the indirect proof from
ref.\cite{FKN91}, deducing (1.29) from eqs.(1.19),(1.20), which are proved in
sect.2,3
without any reference to (1.23)).

If one believes in this transition from (1.26) to (1.29), we can finally return
to multimatrix models. Namely, in \cite{FKN91} it was suggested that
$\sqrt{\Gamma ^{\{K-1\}}_{ds}}$

(i) is a $\tau $-function of $K$-reduced hierarchy,

(ii) satisfies the constraint

\beq
{\cal L}^{\{K\}}_{-1}\sqrt{\Gamma ^{\{K-1\}}_{ds}} = 0\hbox{, }   n \geq  -1,
\eeq
and, as a corollary of (i) and (ii),

(iii) satisfies the entire set of eqs.(1.29):

\beq
{\cal W}^{\{k\}}_{Kn}\sqrt{\Gamma ^{\{K-1\}}_{ds}} = 0\hbox{, }   k =
2,3,...,K\hbox{; }   n \geq  1-k.
\eeq
Now it remains to use the above-mentioned assertion that this system has a
unique solution in order to arrive to our identification

\beq
Z^{\{K\}}_\infty  = \sqrt{\Gamma ^{\{K-1\}}_{ds}}
\eeq
and thus to the central idea of this paper:  {\it universality} of GKM (1.2).

\subsection{Comments}

Before we proceed to actual derivation of all these statements, let us comment
briefly on the very status of the constraints (1.29) imposed on
$\sqrt{\Gamma ^{\{K-1\}}_{ds}}$. In order to honestly derive such constraints
on the lines of \cite{MMMM91} it is first necessary to find out their
counterparts
(loop equations) in the {\it discrete} multimatrix models (these are expressed
in terms of the so called $\tilde {\cal W}$ operators in
\cite{MMM91b,Fut}), then carefully perform the appropriate reductions
(including identification of all the $K-1$ {\it a priori} different
potentials), and take the double-scaling limit. The actual form of the
$\tilde {\cal W}$-constraints is very similar to (1.26) and therefore it may
happen that the final constraints would arise just in the form of (1.26) rather
than (1.29). This could make the relation (1.26) even more important.

After this general description of our reasoning and results it remains to prove
that

(i)  $Z^{\{{{\cal V}}\}}_\infty $ as defined by (1.2)  (the limit $N
\rightarrow
\infty $ is smooth)  is indeed a KP $\tau $-function;

(ii) for any  ${{\cal V}}(X)$  it is subjected to additional constraint (1.25);

(iii) if  ${{\cal V}}(X) = const\cdot X^{K+1}$, then $Z^{\{K\}}_\infty $
in fact becomes
a $\tau $-function $\tau ^{\{K\}}$ of the $K$-reduced KP-hierarchy;

(iv) in this case also the entire set of constraints like (1.15) for $K=2$ and
(1.26) for $K=3$ are imposed on $Z^{\{K\}}_\infty .$

The proof of the points (i),(iii) is given in sect.2, (ii) is discussed in
sect.3
and (iv) --- in sect.4. While (i),(ii),(iii) are proved in full generality ---
for any potential ${{\cal V}}(X)$, --- the universal form of the constraints
(iv)
is not completely understood. Moreover, the proof of (iv) for the case of
${{\cal V}}(X) = \displaystyle{{X^{K+1}\over K+1}}$,
as presented in sect.4, is not complete even
in the simplest case of $K=3$. Also the proof of (iv) is not independent of
(iii), since eq.(1.21) is used. The deep understanding of these
${\cal W}_\infty $-constraints
and their universal formulation remains the first
obvious problem for GKM to be resolved in the future. The second obvious
problem is to find out the algebro-geometrical origin of the model (1.2)
itself.

\section{Partition function of the GKM as KP $\tau $-function}
\setcounter{equation}{0}
\subsection{Short summary}

The purpose of this section is to prove that:

 ($A$) The partition function
$Z^{\{K\}}_N[M]$, if considered as a function of time-variables

\beq
T_n = {1\over n} trM^{-n}\hbox{ ,}
\eeq
is a KP $\tau $-function for {\it any} value of  $N$  and {\it any} potential
${{\cal V}}[X].$

($B$) As soon as  ${{\cal V}}[X]$  is homogeneous polynomial of degree  $K+1$,
$Z^{\{K\}}_N[M]$  is in fact a $\tau $-function of  $K$-reduced KP hierarchy.
Moreover, actually,
$\displaystyle {{\partial Z^{\{K\}}\over \partial T_{nK}}}=0$.

In order to prove these statements, first, in sect. 2.2 we rewrite the $r.h.s$.
of eq.(1.2) in terms of determinant formula (2.21)

\beq
Z^{\{K\}}_N[M] = {{\rm det} _{(ij)}\phi _i(\mu _j)\over \Delta (\mu )}\
\ \ \ \ i.j =
1,...,N.
\eeq
We show in sect. 2.3 that {\it any} KP $\tau $-function in the Miwa
parameterization does have the same determinant form. Finally,
as a check of self-consistency, we prove in
sect. 2.4 that any determinant formula (2.2) with {\it any} set of functions
$\{\phi _i(\mu )\}$  satisfies Hirota difference equation.

A brief description of Grassmannian (or
Universal module space) language is given in sect. 2.5. The point of
Grassmannian corresponding to $\tau $-function (2.2) is defined by the
infinite-vector function
$\{\phi _i(\mu )\}$. Remarkably, the results of ref. \cite{KS91} which concern
identification of the points in Grassmannian, associated with multi-matrix
models, are
immediately reproduced from the explicit expressions for the basis
$\{\phi _i(\mu )\}.$ This exhausts our discussion of ($A$).

The main thing which distinguishes matrix models from the point of view of
solutions to the KP-hierarchy is that the set of functions  $\{\phi _i(\mu )\}$
in (2.2) is not arbitrary. Moreover, this whole {\it infinite} set of functions
is
expressed in terms of a {\it single} potential  ${{\cal V}}[X]$ ($i.e$. instead
of
arbitrary {\it matrix  $A_{ij}$} in  $\phi _i(\mu ) = \sum A_{ij}\mu ^j$
we have here
only a {\it vector}  ${{\cal V}}_i$ or  ${{\cal V}}[\mu ] =
\sum {{\cal V}}_i\mu ^i)$. This is
the origin of  ${\cal L}_{-1}$ and other  ${\cal W}$- constraints (which in the
context of KP-hierarchy may be considered as implications of  ${\cal L}_{-1})$.
All these constraints are in fact
contained in the Ward identity (1.23). While the
detailed discussion of these constraints is postponed to the sect.3,4, in
subsect.2.6
we discuss {\it another} implication of (1.23) arising in the case of
{\it polynomial  ${{\cal V}}[X]$}. Namely, whenever  ${{\cal V}}[X]$  is a
homogeneous
polynomial of degree  $K+1$,  $\phi _{i+K}(\mu )$  in (2.2) can be substituted
by
$\mu ^k\phi _i(\mu )$  and in such points of Grassmannian KP $\tau $-function
is
known to possess the property (1.21) and thus may be considered as a
$\tau $-function  $\tau ^{\{K\}}$ of $K$-reduced KP hierarchy. In sect.2.7 we
prove that the factor (1.22) is actually absent in GKM. This is all to be said
below about ($B$).

\subsection{From GKM to determinant formula}

\bigskip
{\it Numerator of (1.2).}
We begin from evaluation of the integral (1.1):

\beq
{\cal F}^{\{{{\cal V}}\}}_N[\Lambda ] \equiv  \int   dX\ e^{- Tr[{{\cal V}}(X)
-
Tr\Lambda X]}\hbox{.}
\eeq
If eigenvalues of  $X$  and  $\Lambda $  are denoted by  $\{x_i\}$  and
$\{\lambda _i\}$  respectively, this integral can be rewritten as

\beq
{V_N\over \Delta (\Lambda )}\left[  \prod _{i=1}^N
\int   dx_ie^{-{{\cal V}}(x_i)+\lambda _ix_i} \right] \Delta (X)\hbox{.}
\eeq
$V_N$ stands for the volume of unitary group  $U(N)$  and is unessential in
what follows; $\Delta (X)$  and  $\Delta (\Lambda )$  are Van-der-Monde
determinants, $e.g$. $\Delta (X) = \prod_{i>j}(x_i-x_j)$.
The transformation, leading from (2.3) to (2.4), is a trick,
familiar from the study of multi-matrix models in the formalism of orthogonal
polynomials \cite{IZ80,CMM81,Meh81}, and we do not dwell upon this point.
For what follows it is
very important that the term $Tr\Lambda X$ in (2.3) is linear in  $X$  (be it
instead, say,
$Tr\Lambda X^p$ , the polynomial factor  $\Delta (X)$  at the $r.h.s$.
of eq.(2.4) would be substituted by  $\Delta ^2(X)/\Delta (X^p)$  and our
reasoning below would be unapplicable).

The $r.h.s$. of (2.4) is proportional to the determinant of $N\times N$ matrix

\beq
\Delta ^{-1}(\Lambda ) \hbox{det} _{(ij)}F_i(\lambda _j)
\eeq
with

\beq
F_{i+1}(\lambda ) \equiv  \int   dx\ x^ie^{-{{\cal V}}(x)+\lambda x} =
({\partial \over \partial \lambda })^iF_1(\lambda ).
\eeq
Note that

\beq
F_1(\lambda ) = {\cal F}^{\{{{\cal V}}\}}_{N=1}[\lambda ]\hbox{ . }
\eeq
If we recall that in GKM

\beq
\Lambda  = {{\cal V}}'(M)
\eeq
and denote the eigenvalues of  $M$  trough  $\{\mu _i\}$ , eq.(2.6) acquires
the form:

\beq
{\cal F}^{\{{{\cal V}}\}}_N[{{\cal V}}'(M)] = {{\rm det} \ \tilde
\Phi _i(\mu _j)\over
\prod _{i>j}({{\cal V}}'(\mu _i)-{{\cal V}}'(\mu _j))}\ ,
\eeq
where

\beq
\tilde \Phi _i(\mu ) = F_i({{\cal V}}'(\mu )).
\eeq

\bigskip
{\it Denominator.}
Proceed now to the denominator of (1.2):

\beq
I^{\{{{\cal V}}\}}_N[M] \equiv  \int   dX\ e^{-U_2(M,X)}.
\eeq
Making use of $U(N)$-invariance of Haar measure $dX$ one can easily diagonalize
$M$  in (2.11). Of course, this does not imply any integration over angular
variables and provide no factors like  $\Delta (X)$. Then for evaluation of
(2.11) it remains to use the obvious rule of Gaussian integration,

\beq
\int   dX\ e^{-\sum ^N_{i,j} U_{ij}X_{ij}X_{ji}} \sim \prod ^N_{i,j}
U^{-1/2}_{ij}
\eeq
(an unessential constant factor is omitted), and substitute the explicit
expression for $U_{ij}(M)$. If potential is represented as a formal series,

\beq
{{\cal V}}(X) =\sum ^\infty _{n=1}{{\cal V}}_nX^n,
\eeq
(and thus is supposed to be analytic in  $X$  at  $X = 0)$, eq.(1.4) implies
that

$$
U_2(M,X)
=\sum ^\infty _{n=0}(n+1){{\cal V}}_{n+1}\left\lbrace \sum _{a+b=n-1}TrM^aXM^bX
\right\rbrace ,
$$
and

$$
U_{ij} =\sum ^\infty _{n=0}(n+1){{\cal V}}_{n+1} \left\lbrace
\sum _{a+b=n-1}\mu ^a_i\mu ^b_j \right\rbrace  =
\sum ^\infty _{n=0}(n+1){{\cal V}}_{n+1} {\mu ^n_i - \mu ^n_j\over \mu _i -
\mu _j}  = {{{\cal V}}'(\mu _i) - {{\cal V}}'(\mu _j)\over \mu _i - \mu _j}.
$$
Coming back to (1.2), we conclude that

\bea
Z^{\{{{\cal V}}\}}_N[M] = e^{Tr[{{\cal V}}(M)-M{{\cal V}}'(M)]}
{{\cal F}_N[{{\cal V}}'(M)]\over I^{\{K\}}[M]} \sim \nn \\
\sim  [{\rm det} \ \tilde \Phi _i(\mu _j)] \prod _{i>j}^N
{U_{ij}\over ({{\cal V}}'(\mu _i)-{{\cal V}}'(\mu _j))} \prod _{i=1}
s(\mu _i)  =  {[{\rm det} \ \tilde \Phi _i(\mu _j)]\over \Delta (M)}
\prod _{i=1}^N
s(\mu _i)\ .
\eea
Here  $s(\mu _i) \equiv  [U_{ii}]^{1/2}
e^{{{\cal V}}(\mu _i)-\mu _i{{\cal V}}'(\mu _i)}$, $i.e.$

\beq
s(\mu ) = [{{\cal V}}''(\mu )]^{1/2} e^{{{\cal V}}(\mu )-\mu {{\cal V}}'(\mu )}
\eeq
The product of $s$-factors at the $r.h.s$. of (2.16) can be absorbed into
$\tilde \Phi $-functions:

\beq
Z^{\{{{\cal V}}\}}_N[M] = {[{\rm det}
\Phi _i(\mu _j)]\over \Delta (M)}\hbox{,}
\eeq
where

\beq
\Phi _i(\mu ) = s(\mu )\tilde \Phi _i(\mu ).
\eeq

\bigskip
{\it Kac-Schwarz operator.}
{}From eqs.(2.6),(2.10) and (2.15) one can deduce that  $\Phi _i(\mu )$
can be derived from the basic function  $\Phi _1(\mu )$  by the relation

\beq
\Phi _i(\mu ) = [{{\cal V}}''(\mu )]^{1/2}\int   (y +
\mu )^{i-1}e^{-U(\mu ,y)}dy =
A_{\{{\cal V}\}}^{i-1}(\mu )\Phi _1(\mu )\ ,
\eeq
where  $U(\mu ,y)$  is defined by eq.(1.3) and  $A_{\{{\cal V}\}}(\mu )$  is
the
first-order
differential operator

\bea
A_{\{{\cal V}\}}(\mu ) & = & s {\partial \over \partial \lambda } s^{-1} =
{e^{{{\cal V}}(\mu )-\mu {{\cal V}}'(m)}\over [{{\cal V}}''(\mu )]^{1/2}}
{\partial \over \partial \mu }
{e^{-{{\cal V}}(\mu )+\mu {{\cal V}}'(\mu )}\over [{{\cal V}}''(\mu )]^{1/2}} =
\nn \\
& = & {1\over {{\cal V}}''(\mu )} {\partial \over \partial \mu } + \mu  -
{{{\cal V}}'''(\mu )\over 2[{{\cal V}}''(\mu )]^2}\ .
\eea
In the particular case (1.9)

$$
A_{\{K\}}(\mu ) =
{1\over k\mu ^{k-1}} {\partial \over \partial \mu } + \mu  -
{k-1\over 2k\mu ^k}
$$
coincides (up to the scale transformation of  $\mu $  and
$A_{\{K\}}(\mu ) )$ with the operator which determines the finite dimensional
subspace
of the Grassmannian in ref.\cite{KS91} (we shall return to this point in
sect.2.5).
We emphisize that the property

\beq
\Phi _{i+1}(\mu ) = A_{\{{\cal V}\}}(\mu ) \Phi _i(\mu )\ \ \ \
( F _{i+1}(\lambda )
= {\partial \over \partial \lambda } F _i(\lambda ) )
\eeq
is exactly the thing which distinguishes partition functions of GKM from
expression (2.2) for generic $\tau $-function,

\beq
\tau ^{\{ \phi _i\}}_N[M] = {[{\rm det} \ \phi _i(\mu _j)]\over \Delta (M)}
\hbox{,}
\eeq
with arbitrary sets of functions  $\phi _i(\mu )$. In the next section we
demonstrate that the quantity (2.21) is in fact a KP $\tau $-function in Miwa
coordinates. Implications of additional constraint (2.20) will be discussed in
sect.2.6 and sect.3,4.

\bigskip
{\it $N$-dependence.}
Before we proceed to discussion of Miwa coordinates, two more things about the
formula (2.21) deserve mentioning. First, the entire set
$\{\phi _i(\mu )\}$  is
naturally  $N$-independent and infinite. It is reasonable to require also that
$\phi _i$'s  are linear independent and

\beq
\phi _i(\mu )/\phi _1(\mu ) = c_i\mu ^{i-1}(1 + o({1\over \mu }))\hbox{,}
\eeq
with  $\mu $-independent  $c_i$ (this is true for functions (2.7)). Then the
set  $\{\phi _i(\mu )\}$  is identified as projective coordinates of a point of
Grassmannian, see sect.2.5. The $r.h.s$. of eq.(2.21) naturally represents
$\tau ^{\{\phi _i\}}_\infty [M]$  for an infinitely large matrix  $M$. In order
to return to the case of finite  $N$  it is enough to require that all
eigenvalues of  $M$ , except  $\mu _1,\ldots,\mu _N$, tend to infinity. In this
sense the function  $\tau ^{\{\phi _i\}}_N[M]$  in (2.21) is independent of
$N$ ; the entire dependence on  $N$  comes from the argument  $M$: $N$  is the
quantity of finite eigenvalues of  $M$ . As a simple check of consistency, let
us additionally carry  $\mu _N$ to infinity in (2.21), then, according to
(2.22)

$$
\hbox{det} _N\phi _i(\mu _j) \sim
c_N\phi _1(\mu _N)(\mu _N)^{N-1}\cdot \hbox{det} _{N-1}
\phi _i(\mu _j)\cdot (1 +
o(1/\mu _N))
$$
and

$$
\Delta _N(M) \sim  (\mu _N)^{N-1}\Delta _{N-1}(M)(1 + o(1/\mu _N)),
$$
so that

\beq
\tau ^{\{\phi _i\}}_{N _{\mu _N}} \stackreb{\mu _N \to \infty}{\sim}
\tau ^{\{\phi _i\}}_{N-1}\cdot [c_N\phi _1(\mu _N)](1+o
(1/\mu _N))\hbox{.}
\eeq
This is the exact statement about the $N$-dependence of  $\tau _N$. Actually in
GKM  $c_N\phi _1(\mu ) = 1$  at  $\mu  \rightarrow  \infty $ . In what follows
we often omit the subscript $N.$

\bigskip
{\it Multiplicities.}
The second remark concerns the form of the $r.h.s$. of (2.21), when some
eigenvalues of  $M$  coincide (see also eq.(2.30) below). If eigenvalue
$\mu _i$ appears with the multiplicity $p_i$ , eq.(2.21) looks like

\beq
\tau [(\mu _i,p_i)] \sim  {{\rm det}  [\phi _i(\mu _j) A_{\{{\cal V}\}}
\phi _i(\mu _j)\hbox{ ... }
A_{\{{\cal V}\}}^{p_j-1}\phi _i(\mu _j)]\over \prod _{i>j}(\mu _i-
\mu _j)^{p_ip_j}}.
\eeq
Notation in this formula is not very transparent: it is implied that the matrix
in the numerator has rows of the form

$$
\new
\begin{array}{l}
\phi _i(\mu _1),\ A_{\{{\cal V}\}}\phi _i(\mu _1),\hbox{ ... },
A_{\{{\cal V}\}}
^{p_1-1}\phi _i(\mu _1),  \\
\ \phi _i(\mu _2),\   A_{\{{\cal V}\}}\phi _i(\mu _2),
\hbox{ ... }, A_{\{{\cal V}\}}^{p_2-1}\phi _i(\mu _2),\ \phi _i(\mu _3),\
A_{\{{\cal V}\}}\phi _i(\mu _3),\hbox{ ... },
\end{array}
$$
where operator $A_{\{{\cal V}\}}$  is defined by eq.(2.19). Expression (2.24)
will be used in
sect.2.4 in the proof of Hirota identity. Note also that if some  $\mu _i
\rightarrow  \infty $, this may be treated as vanishing of the corresponding
multiplicity,  $p_i = 0$. Such  $\mu _i$ obviously do not contribute to (2.24).
Thus, the value of $N$ (the finite size of the matrix  $M$  in GKM)
may be interpreted as the number of  $\mu$'s  which
appear with non-vanishing multiplicities.

\subsection{KP $\tau $-function in Miwa parameterization}

Generic KP $\tau $-function is a correlator of a special form:

\beq
\tau ^G\{T_n\} = \langle 0|:e^{\sum \ T_nJ_n}: G|0\rangle
\eeq
with

\beq
J(z) = \tilde \psi (z)\psi (z)\hbox{; }  G =\
:\exp \ {\cal G}_{mn}\tilde \psi _m\psi _n:
\eeq
in the theory of free 2-dimensional fermionic fields  $\psi (z)$,
$\tilde \psi (z)$  with the action $\int
\tilde \psi \bar \partial \psi $. Vacuum states are defined by conditions

\beq
\psi _n|0\rangle  = 0\ \ n < 0\hbox{ , }  \tilde \psi _n|0\rangle  = 0\ \ n
\geq  0
\eeq
where  $\psi (z) =
\sum _{\bf Z}
\psi _nz^n\ dz^{1/2} $ , $\tilde \psi (z) =
\sum _{\bf Z}
\tilde \psi _nz^{-n-1}\ dz^{1/2}$.

The crucial restriction on the form of the correlator, implied by (2.26) is
that the operator  $:e^{\sum \ T_nJ_n}:$  $G$  is Gaussian exponential
(exponent is
quadratic in fields), so that insertion of this operator may be considered as
modification of  $\tilde \psi -\psi $  {\it propagator}, and Wick's theorem is
usually applicable. Namely, the correlators

\beq
C_G(\{\mu _i\},\{\lambda _j\}) \equiv  \langle 0|
\prod _i
\tilde \psi (\mu _i)\psi (\lambda _i) G|0\rangle
\eeq
for {\it any} relevant $G$ are expressed through the pair correlators of the
same form:

\beq
C_G(\{\mu _i\},\{\lambda _j\}) = {\rm det} _{(ij)} C_G\{\mu _i,\lambda _j\}.
\eeq
Operator  $G$  in (2.28) can be safely substituted by the entire
$:e^{\sum T_nJ_n}:G$ , but we do not need this for our purposes below.
Instead with the help of Miwa transformation we shall express
$:e^{\sum T_nJ_n}:$  in (2.25) through insertions of fermionic operators.
Then (2.25) acquires the form of (2.28) and after that the application (2.29)
provides a representation for $\tau $-function in determinant form, to be
compared with eq.(2.21) for partition function of GKM.

There are slightly different forms of Miwa transformation. Usually one writes

\beq
T_n = {1\over n} \sum _i p_i{1\over \mu ^n_{_i}}
\eeq
and treats the $r.h.s$. as an integral  $\displaystyle {\int
{p(\mu )\over \mu ^n}}$  over
Riemann sphere with coordinate  $\mu $  and singular function  $p(\mu )
= \sum _i p_i\delta (\mu -\mu _i)$. We use instead the transformation (2.1):

\beq
T_n = {1\over n} Tr\ M^{-n} = {1\over n} \sum _i {1\over \mu ^n_i}\hbox{ ,}
\eeq
interpreting  $\mu _i$'s  as eigenvalues of the matrix  $M$ , all coming with
unit multiplicities $p_i= 1$ (while $p_i > 1$ in (2.30) may be interpreted as
result of coincidence of $p_i$ eigenvalues). The form (2.30) of Miwa
transformation is preferable from the point of view of invertibility and also
is convenient for the proof of Hirota identities (see sect.2.4 below).

The simplest way to understand what happens to the operator  $e^{\sum T_nJ_n}$
after the substitution of (2.31), is to use the free-{\it boson} representation
of the current  $J(z)=\partial \varphi (z)$. Then  $\sum  T_nJ_n =
\displaystyle {\sum _i
\left\lbrace  \sum _n {1\over n\cdot \mu _i^n}
\varphi _n\right\rbrace}  = \sum _i \varphi (\mu _i)$, and

\beq
:e^{\sum _i\varphi (\mu _i)}: = {1\over \prod _{i<j}(\mu _i-\mu _j)}
\prod _i :e^{\varphi (\mu _i)}:\hbox{ .}
\eeq
In fermionic representation it is better to start from

\beq
T_n = {1\over n} \sum _i ({1\over \mu ^n_i} - {1\over \tilde \mu _i^n} )
\eeq
instead of (2.31). Then

\beq
:e^{\sum T_nJ_n}: = {\prod _{i,j}^N(\tilde \mu _i-\mu _j)\over
\prod _{i>j}(\mu _i-\mu _j) \prod _{i>j}(\tilde \mu _i-\tilde \mu _j)}
\prod _i \tilde \psi (\tilde \mu _i)\psi (\mu _i)\hbox{ .}
\eeq
In order to come back to (2.31) and (2.32) it is necessary to shift all
$\tilde \mu _i$'s  to infinity. This may be expressed by saying that the left
vacuum in (2.25) is substituted by

$$
\langle N| \sim
\langle 0|\tilde \psi (\infty )\tilde \psi '(\infty )...\tilde \psi ^{(N-1)}(
\infty ).
$$
The $\tau $-function (2.25) can be represented in various forms:

\beq
\new
\begin{array}{l}
\tau ^G_N[M] = \langle 0|:e^{\sum T_nJ_n}:G|0\rangle  =
\Delta (M)^{-1}\langle N|\prod _i
:e^{\varphi (\mu _i)}: G|0\rangle  = \\
=\lim _{\tilde \mu _j \to \infty}{\prod _{i,j}(\tilde \mu _i-\mu _j)
\over \prod _{i>j}
(\mu _i-\mu _j) \prod _{i>j}(\tilde \mu _i-\tilde \mu _j)}\langle 0|\prod _i
\tilde \psi (\tilde \mu _i)\psi (\mu _i)G|0\rangle  =
\lim _{\tilde \mu _j\rightarrow \infty}
{C_G(\{\tilde \mu _i\},\{\mu _i\})\over C_I(\{\tilde \mu _i\},\{\mu _i\})},
\end{array}
\eeq
and applying the Wick's theorem (2.29) we obtain:

\beq
\tau ^G_N[M] = \lim _{\tilde \mu _j\rightarrow \infty }\hbox{det} _{(ij)}
{C_G(\tilde \mu _i,\mu _j)\over C_I(\tilde \mu _i,\mu _j)} =
{{\rm det} \ \phi _i(\mu _j)\over \Delta (M)}
\eeq
with functions

\beq
\phi _i(\mu ) \sim  \langle 0|\tilde \psi ^{(i-1)}(\infty )\psi (\mu )
G|0\rangle .
\eeq
Thus, we proved that KP $\tau $-function in Miwa coordinates (2.2) has the
determinant form (2.1).

Below we shall need a more detailed expression of the right hand side in the
eq.(2.35), when all points  $\tilde \mu _i$ tend to the same value  $1/t$  with
fixed  $t \ne 0$. From (2.35) one can obtain:

\bea
\tau ^G_N[t|M] \equiv {\prod _i (1-t\mu _i)^N\over \prod _{i>j}
(\mu _i-\mu _j)} {\rm det}
\left\lbrace {1\over (i-1)!}\partial ^{i-1}_t\langle 0|\tilde \psi (t^{-1})
\psi (\mu _j) G|0\rangle \right\rbrace  = \nn \\
=  {\prod _i (1-t\mu _i)^N\over \prod _{i>j}
(\mu _i-\mu _j)} {\rm det}  \left\lbrace {1\over (i-1)!} \partial ^{i-1}_t
{\langle 0|:e^{\Sigma {1\over n}(\mu ^{-n}_j-t^n)J_n}:G|0\rangle \over 1 -
t\mu _j}\right\rbrace
\eea
In the limit of $t \to 0$ we obtain the formula (2.37) with

\beq
\phi _i(\mu ) = \lim _{t \to 0} \phi _i(\mu ,t)=
{1\over (i-1)!} \lim _{t \to 0} \partial ^{i-1}_t
{\langle 0|:e^{\Sigma {1\over n}(\mu ^{-n}-t^n)J_n}:G|0\rangle \over 1 -
t\mu }\ .
\eeq
Eqs.(2.38) and (2.39) will be used in sect.3.2 in one of our derivations of the
${\cal L}_{-1}$- constraint.

One more remark is in order now. From the form of the operator expansion for
$\tilde \psi (\infty )\psi (\mu )$  it is clear that the function
$\phi _i(\mu )$ ,  as given by (2.39), behaves as
$\mu ^{i-1}(1 + o(1/\mu ))$ when $\mu  \rightarrow  \infty $, in
accordance with (2.37), and all the functions $\phi _i(\mu )$ are independent.
This last statement should be more or less clear from the fact, that the set
$\{\phi _i(\mu )\}$  contains all the information about fermionic propagator
$C_G(\tilde \mu ,\mu )$ , which is exactly the same as the information in
operator  $G$ , which should be of the form (2.26), and defines the actual
fermionic action. To make this mutual independence of  $\phi$'s  even more
transparent we prove in the next section 2.4 that Hirota identities are
satisfied by the quantity (2.36) with {\it any} set of functions
$\{\phi _i(\mu )\}$. The possibility to choose $\{\phi _i(\mu )\}$  in
{\it any} way is important for us to argue without going into any more details,
that partition function of GKM, which was expressed in determinant form (2.36)
with specific choice of  $\phi _i = \Phi _i$ , is indeed a KP
$\tau $-function.

\subsection{Hirota equations for $\tau $-function in Miwa coordinates}

The usual form of bilinear Hirota equations, which are the defining equations
of KP $\tau $-functions, is:

$$
\sum ^\infty _{i=0} {\cal P}_i(-2y){\cal P}_{i+1}(\tilde D_T)
e^{[\sum _iy_iD_{T_i}]} \tau \cdot \tau  = 0
$$
where  $D_T$ are Hirota symbols, $\tilde D \equiv  (D_{T_{^1}}$, ${1\over 2}
D_{T_{^2}},\ldots$) and  ${\cal P}_i$ are Schur polynomials.

Note that these equations are in fact more than KP equations themselves, which
describe evolution of the functions  $u_i(T_n)$ ($u(T_n) \equiv u_2(T_n) \equiv
\displaystyle
{{\partial ^2\log \ \tau \over \partial T^2_1}}$ and all other $u_i(T_n)$ can
be
determined from the relations $\displaystyle {{\partial \over \partial T_1}{
\partial \over \partial T_n}}\log \tau = (L^n)_{-1}$ with $L \equiv \partial +
\sum _i^\infty u_{i+1}\partial ^{-i}$ and
$\partial \equiv \displaystyle {{\partial
\over \partial T_1}}$ --- see (3.35) and (3.36)) rather than $\tau$-itself.
Indeed, any transformation

\beq
\tau (T_n) \rightarrow  \tau (T_n) e^{H(T_2,T_3...) +
T_1 \tilde H}
\eeq
with  $\partial H/\partial T_1 = 0$  and $\tilde H = const$ does not
change  $u_i(T_n)$, $i.e$. $H$ and $\tilde H$ can not be fixed by the entire
set
of KP equations. They are, however, fixed by Hirota equations (up to linear
function  $H + T_1\tilde H = \sum \ b_nT_n$,  $b_n= const)$. This remark is
important for the subject of matrix models when ever
interpolation between critical
points is considered.  For example, one can use the representation

\beq
\tau (T_n) = \exp  \{\int  ^{T_1}
(T_1-x)u(x,T_2\ldots)dx + H(T_2,T_3\ldots) + T_1\tilde H\}
\eeq
and at any given multicritical point the functions  $H$, $\tilde H$  are
unessential. However, they are very important for interpolations, and, in
particular, for the relevance of Virasoro and  ${\cal W}$-constraints.

Our first task in this section is to rewrite the Hirota equations in Miwa
coordinates (2.1). Since this is a widely known transformation (see, for
example \cite{Miw82}), we just cite here the answer: Hirota equations in Miwa
coordinates state that

\bea
(\mu _a- \mu _b)\tau (p_a, p_b, p_c+1) \tau (p_a+1, p_b+1, p_c) +\nn \\
+ (\mu _b- \mu _c)\tau (p_a+1, p_b, p_c) \tau (p_a, p_b+1, p_c+1)
+ \nn \\
+ (\mu _c- \mu _a)\tau (p_a, p_b+1, p_c) \tau (p_a+1, p_b, p_c+1) =
0,
\eea
where $\tau $-function is expressed through the Miwa variables (2.30) according
to eq.(2.24) and three arbitrary eigenvalues from the set  $\{\mu _i\}$
with corresponding multiplicities are chosen.

The second task is to prove that  $\tau [M] = \Delta (M)^{-1}
{\rm det} \ \phi _i(\mu _j)$  with any $\{\phi _i(\mu )\}$ satisfies (2.42).

All we need to derive the desired equations is the well known Jacobi identity
for the determinants. For any  $(N+2)\times (N+2)$  determinant  $J$  we denote
by  $J\left( {i_1\ldots i_s}\atop {j_1 \ldots j_s} \right) $ the determinant
obtained from  $J$  with  $s$  rows  $i_1$, ... ,
$i_s$ and  $s$  columns  $j_1,\ldots, j_s$ omitted. Then the Jacobi identity
reads

$$
J\left( {i,j}\atop {i,j}\right) J  = J\left({i}\atop {i}\right)
J\left( {j}\atop {j}\right)  -
J\left( {i}\atop {j}\right) J\left( {j}\atop {i}\right)\ .
$$
Let us consider  $\tau _N(\{\mu \}) \equiv  \tau _N(\mu _1,\ldots, \mu _N)$
and divide the given ( {\it a priori} different) set of eigenvalues  $\mu _1,
\ldots, \mu _N$ into the  $L$  ``clusters" of sizes  $p_1,\ldots, p_L :$

$$
\mu _1\hbox{, ... , } \mu _{p_1}\hbox{; } \mu _{p_1+1}\hbox{, ... , }
\mu _{p_1+p_2}\hbox{; ... } ;\mu _{p_1+...+p_{L-1}+1}\hbox{, ... , }
\mu _{p_1+...+p_L}\ ,
$$
where  $\sum _j
p_j = N$. Then one should introduce two additional eigenvalues  $\mu _{N+1}$,
$\mu _{N+2}$ and apply the Jacobi identity for  $i = N+1$, $j = N+2$
to the function
$\tau _{N+2}(\{\mu \}$, $\mu _{N+1}$, $\mu _{N+2})$. Then simple calculation
gives the following system of equations:

\beq
\new
\begin{array}{c}
\tau _{N+2}(\{\mu \}\hbox{, } \mu _{N+1}\hbox{, } \mu _{N+2}) \tau _N(\{\mu \})
 =  \\
 =  {1\over \mu _{N+1}- \mu _{N+2}} \left\lbrace \tau _{N+1}(\{\mu \}\hbox{, }
\mu _{N+2})\hat \tau _{N+1}(\{\mu \}\hbox{, } \mu _{N+1})\right. - \\
 -
\left.\tau _{N+1}(\{\mu \}\hbox{, } \mu _{N+1})\hat \tau _{N+1}(\{\mu \}
\hbox{, }
\mu _{N+2})\right\rbrace
\end{array}
\eeq
where  $\hat \tau _{N+1}$ denotes some new $\tau $-function which is obtained
from the given  $\tau _{N+1}$ by  a change of the last row:  $\phi _{N+1}
\rightarrow  \phi _{N+2}$ .  Now let all the eigenvalues in each cluster
tend to the values  $\mu _1 ,\ldots, \mu _L$ respectively and
$\mu _{N+1}\rightarrow  \mu _a$, $\mu _{N+2}\rightarrow  \mu _b$ where
$\mu _a$ and  $\mu _b$ belong to different clusters. Then eq.(2.43) acquires
the form (in the notations as in eq.(2.42)):

\beq
\new
\begin{array}{c}
(\mu _a - \mu _b)\tau (p_a + 1\hbox{, } p_b + 1\hbox{, } p_c)
\tau (p_a\hbox{, } p_b\hbox{, } p_c) = \\
= \tau (p_a\hbox{, } p_b + 1\hbox{, } p_c)\hat \tau (p_a + 1\hbox{, }
p_b\hbox{, } p_c) -
\tau (p_a + 1\hbox{, } p_b\hbox{, } p_c)\hat \tau (p_a\hbox{, } p_b +
1\hbox{, } p_c)\ ,
\end{array}
\eeq
where  $p_c$ describes some arbitrary third cluster different from the previous
ones. One can multiply this equation by the factor
$\displaystyle {{\tau (p_a,p_b,p_c+
1)\over \tau (p_a,p_b,p_c)}}$  and write down the couple of two another
equations obtained by cyclic permutations among indices ($a$, $b$, $c)$. The
sum
of these three equations coincides with eq.(2.42).

Another interesting expression can be derived from the eq.(2.44) as follows.
Let
us make the shift  $p_a \rightarrow  p_a - 1$, $N \rightarrow  N-1$ and put
$p_b = 0$ (this means that we have the cluster with the single element
$\mu _b$ in the corresponding $\tau $-functions in eq.(2.44)). Then in the
limit
$\mu _b \rightarrow  \mu _a$ we have  $\tau (p_a$, $p_b + 1) \rightarrow
\tau (p_a+1)$ (notations of other clusters will be omitted since they don't
changed) and equation now takes the form

$$
\tau _{N+1}(p_a + 1) \tau _{N-1}(p_a - 1) =
$$

$$
= \tau _N(p_a) \lim _{\mu _b \to \mu _a}
{\partial \over \partial \mu _b}\hat \tau _N(p_a\hbox{, }
\mu _b) - \hat \tau _N(p_a) \lim _{\mu _b \to \mu _a}
{\partial \over \partial \mu _b}\tau _N(p_a\hbox{, } \mu _b)\hbox{  .}
$$
Simple calculation gives further that

$$
\lim _{\mu _b \to \mu _a}
{\partial \over \partial \mu _b}\tau _N(p_a,\ \mu _b) = {1\over p_a}
{\partial \over \partial \mu _a}\tau _N(p_a)
$$
and now we obtain

\beq
p_a \tau _{N+1}(p_a + 1) \tau _{N-1}(p_a - 1)  = \tau ^2_N(p_a)
{\partial \over \partial \mu _a}{\hat \tau (p_a)\over \tau (p_a)}\ .
\eeq
This equation holds for every function in the form (2.21). More concrete
expression can be obtained for the GKM partition function
 ${\cal F}^{\{{{\cal V}}\}}_N$
defined by eq.(2.9) in the terms of  $\lambda $'s  variables (see eq.(2.8)).
For this quantity the following relation holds:

$$
\hat {\cal F}^{\{{{\cal V}}\}}_N = \sum ^L_{c=1}
{\partial \over \partial \lambda _c}{\cal F}^{\{{{\cal V}}\}}_N\hbox{ . }
$$
Therefore we have

$$
p_a {\cal F}^{\{{{\cal V}}\}}_{N+1}(p_a + 1) {\cal F}^{\{{{\cal V}}\}}_{N-1}
(p_a -
1) =
$$

$$
= {\cal F}^{\{{{\cal V}}\}}_N(p_a)
\sum ^L_{b=1}{\partial ^2\over \partial \lambda _a\partial \lambda _b}{\cal F}^
{\{{{\cal V}}\}}_N(p_a) -
{\partial \over \partial \lambda _a}{\cal F}^{\{{{\cal V}}\}}_N(p_a)
 \sum ^L_{b=1}
{\partial \over \partial \lambda _b}{\cal F}^{\{{{\cal V}}\}}_N(p_a)\hbox{ .}
$$
We should remark that in the case of  $p_1 = N$,  $p_a = 0,\ \ a \geq  1$ ($
\lambda _i \equiv  \lambda  )$ this equation reduces to the Toda-chain one.
Indeed, now  ${\cal F}^{\{{{\cal V}}\}}_N(p_a)$  has very simple determinant
form

$$
{\cal F}^{\{{{\cal V}}\}}_N(\lambda ) = {1\over (N-1)!}
{\rm det} \ \partial ^{i+j}F_1(\lambda )\ .
$$
It is just determinant of matrix with specific Toda-chain symmetry of constant
entries along anti-diagonal \cite{KMMOZ91}. From the other
point of view, matrix integral (1.1) reduces in the present case (of
proportional to unit
matrix $\Lambda$) to standard discrete model partition function
which is well-known to correspond to Toda-chain hierarchy
\cite{GMMMO91,GMMMMO91}.
If we return to
the original description in the terms of  $\mu $'s  (see eqs.(2.16), (2.24)),
then our final equation is

\bea
p_a{\tau _{N+1}(p_a+1)\tau _{N-1}(p_a-1)\over {{\cal V}}''(\mu _a)} =
\tau ^2(p_a)
\sum ^L_{b=1}{\partial ^2\over \partial \lambda _a\partial \lambda _b}\log
{\tau _N(p_a)\over {\cal N}}\ ,\\
{\cal N} = \left( \prod _{m>n}(\mu _m -
\mu _n)^{p_mp_n}\right) ^{-1}\prod ^L_{n=1}\left( [{{\cal V}}''(\mu _n)]^{1/2}
e^{{{\cal V}}(\mu _n)-\mu _n{{\cal V}}'(\mu _n)}\right) ^{p_n}\ .\nn
\eea
We shall return to eq.(2.46) in sect.3.3 in the context of the string equation.

\subsection{Grassmannian description of KP $\tau $-function}

In this section we present the necessary material about infinite dimensional
Grassmannians and their connection to integrable hierarchies. The details and
proofs can be found in \cite{Sat81,SW85,Zab89}.

\bigskip
{\it Definitions.}
Let us consider the infinite dimensional Hilbert space $H$ realized as the
space of Laurent series  $\phi (\mu )$  on the unit circle  $S^1 (\mu  \in
S^1)$  on the complex plane. This space is naturally represented as the direct
sum  $H = H_+\oplus H_-$ , where  $H_+ (H_-)$ is spanned by the vectors
$\{\mu ^n\}$, $n \geq  0$ ($\{\mu ^{-n}\}$, $n > 0)$, $n \in  {\bf Z}$. These
vectors form the orthogonal basis in  $H$: $\langle \mu ^n$, $\mu ^m\rangle  =
\delta _{nm}.$

The infinite dimensional Grassmannian  $Gr$  is the set of subspaces
$W \in  H$ (Grassmannian point) satisfy the two conditions:

1) the orthogonal projector  $pr_+:W \rightarrow  H_+$ is Fredholm operator,
$i.e$. the kernel and co-kernel are finite dimensional spaces;

2) The orthogonal projector  $pr_-:W \rightarrow  H_-$ is compact operator.

This definition corresponds to Sato's Grassmannian rather than Segal-Wilson
one, where $\phi (\mu )$ should be convergent in some neighborhood of
$\infty $ .

The index of  $pr_+$ is called the relative dimension of  $W:$

$$
\hbox{ind} \ pr_+ = \dim(ker \ pr_+)\  - \ \dim(coker \ pr_+).
$$
Grassmannian is decomposed into the direct sum of the connected components:

\beq
Gr = \stackreb{n}{\oplus} Gr_n,
\eeq
where  $Gr_n$ consists of  $W$'s  with the relative dimension  $n$. For our
purposes it is sufficient to deal with  $Gr_0.$

It is convenient to represent linear operators acting in $H$ as block matrices
with respect to the decomposition  $H_+\oplus H_-:$

\beq
\alpha  = \left[
\begin{array}{cc}
a & b \\ c & d \\ \end{array} \right] \hbox{ , } \alpha  \in GL(\infty ),
\eeq
where the operators  $a$, $b$, $c$, $d$ act as follows: $a$:
$H_+\rightarrow H_+$, $b$: $H_-\rightarrow H_+$, and so on. Clearly, $b$ and
$c$
should be compact operators.

For applications to integrable hierarchies the following commutative subgroup
$\Gamma \subset GL(\infty )$
is of importance. It consists of the mappings from  $S^1$
to non-zero complex numbers and acts on  $H$  by multiplications of the Laurent
series. There are two natural commutative subgroups  $\Gamma _+$ and
$\Gamma _-$ in  $\Gamma :$ the elements of  $\Gamma _+ (\Gamma _-)$  can be
analytically continued inside (outside)  $S^1$. Their elements
$\gamma _+(\mu ) \in  \Gamma _+$ and  $\gamma _-(\mu ) \in  \Gamma _-$ can be
parameterized through the infinite sets  $\{T_k\}$  and  $\{\tilde T_k\}$ ($k
\geq  1)$ of independent variables (``times"):

\beq
\gamma _+(\mu ) = \exp  (\sum _{k\geq 1}T_k\mu ^k)\hbox{, } \gamma _-(\mu ) =
\exp  (\sum _{k\geq 1}\tilde T_k\mu ^{-k}).
\eeq
Let  $\{\phi _n\}$  $(n \geq  0)$ be a basis in  $W$. It is called
{\it admissible}
if the operator transforming  $\mu ^n$ to  $pr_+(\phi _n)$  in  $H_+$ has a
well-defined determinant.

We can write

\beq
\phi _n = \sum ^\infty _{k=1} (\phi _+)_{nk}\mu ^{k-1} +
\sum ^\infty _{k=1}(\phi _-)_{nk}\mu ^{-k},
\eeq
where  $\phi _+$ has a determinant,  $\phi _-$ is compact. In particular, for
$W \in  Gr_0$  it is always possible to choose  $(\phi ^{(c)}_+)_{mn} =
\delta _{mn}$. In this case we call this basis $\{\phi ^{(c)}\}$
{\it canonical} for
$W$. Generally, the convenient choice for  admissible $\phi _+$
(to be used below) is a lower-triangular
matrix with unit diagonal elements: $(\phi _+)_{nk} = 0$ if $n < k$. In this
case

\beq
\phi _k(\mu ) = \mu ^{k-1}( 1+ O(\mu ^{-1}))\hbox{, } k\geq 1
\eeq
(compare to (2.22)).

The $\tau $-function of KP hierarchy  $\tau _W$ is defined as determinant of
the orthogonal projection  $\gamma _+W\rightarrow H_+:$

\beq
\tau _W(\{T_k\}) = {\rm det} (pr_+: \gamma _+(\{T_k\})W\rightarrow H_+),
\eeq
or, more explicitly,

\beq
\tau _W(\{T_k\}) = {\rm det} (1+a^{-1}bB),
\eeq
where  $\gamma _+$ is represented in the block form (2.48),  $B =
\phi _-(\phi _+)^{-1}$ (see (2.50)).

Given a Riemann surface of finite genus one can associate with it a point in
Gr  by constructing the admissible basis according to the well-known rules
\cite{SW85}. In this case the formal Laurent series for the basis vectors are
in fact
convergent. Our situation is different: our basis vectors are asymptotic
(formal) series which do not need to
converge. This should be interpreted as adequate
generalization to infinite genus Riemann surfaces. In this sense we often
identify $Gr$ with the universal module space (of line bundles over Riemann
surfaces with punctures).

Instead of  $W$  one can work with the equivalent space
$\gamma _-(\{\tilde T_k\})W$, the $\tau $-function being changed in the trivial
way:

\beq
\tau _{\gamma _-W}(\{T_k\}) = \exp (\sum ^\infty _{k=1}kT_k\tilde T_k)
\tau _W(\{T_k\})\ .
\eeq

Let us note that in Miwa variables (2.30) the ``evolution
factor"  $\gamma _+(\mu )$  $(2.49)$ looks as a product of pole factors:

$$
\gamma _+(\mu ) = \prod  _j (1 - {\mu \over \mu _j})^{-p_j}\ .
$$

The various reductions of the KP hierarchy can be described in these terms as
follows. Let  $f(\mu )$  be a Laurent series in  $\mu $. Those  $W$'s  which
satisfy the condition

\beq
f(\mu )W \subset  W
\eeq
give rise to solutions of the $``f$-reduction" of KP. In particular, the
choice  $f(\mu ) = \mu ^K$ corresponds to the well known $K$-reductions (KdV,
Boussinesq, ...) which are in fact associated with  $A_{K-1}$ Lie algebra
\cite{DrS84}.
In this case the $\tau $-function does not depend on $T_n$ with  $n = 0$
{\it mod K} (modulo trivial exponential factors linear in times like that
in (2.54)). An admissible (but non-canonical)
basis can be chosen in such a way that

\beq
\phi _{K+n}(\mu ) = \mu ^K\phi _n(\mu )\hbox{, }    n \geq  1.
\eeq
Vise versa, if we have a basis of the form (2.56), the corresponding point of
$Gr$ leads to a solution of the $K$-reduced KP hierarchy. In the next
section 2.6 we show that the GKM (1.2) with arbitrary {\it polynomial}
potential  ${{\cal V}}(x)$  corresponds to the  ${{\cal V}}'(\mu )$-reduction
of the KP hierarchy.

Note that $K$-reduction is equivalent to existance of {\it some}
admissible basis with the property (2.56). However, other admissible
basises (including canonical) in this case satisfy a weaker condition:

$$
\phi _{K+n}(\mu ) = \mu ^K \cdot \sum _{i=1}^n a_i\phi _i
$$
with some constant $a_i$'s.

\bigskip
{\it Fermionic formalism.}
Grassmannian approach is in fact equivalent to the fermionic language used in
the section 2.3. Let us briefly describe the relation between them.

In the fermionic interpretation  $H$  becomes the one-particle Fock space
spanned by $\tilde \psi _n$. The fermionic operator  $G$  of the form (2.26)
makes linear transformations in  $H$ according to

$$
\tilde \psi _n\rightarrow  G\tilde \psi _nG^{-1} \in  H.
$$
The positive (negative) modes of the current  $J(\mu )$  $(2.26)$ are the
generators of $\Gamma _+(\Gamma _-)$. The matrix  ${\cal G}_{mn}$ in (2.26) can
be expressed through the {\it canonical} basis  $\phi _k^{(c)}(\mu )$ of the
corresponding  $W$  as follows:

\beq
\delta _{mn} + G_{mn} = {1\over (2\pi i)^2}\oint
{d\mu d\nu \over \mu \nu } \mu ^m\nu ^n {\cal G}_W(\mu ,\nu ),
\eeq

\beq
{\cal G}_W(\mu ,\nu ) = \sum ^\infty _{k=1}\mu ^{-k}\phi _k^{(c)}(\nu ).
\eeq
A more invariant expression is

$$
{\cal G}_W(\mu ,\nu ) = {1\over \mu -\nu }
\tau _W(\{{\nu ^{-k}-\mu ^{-k}\over k}\}).
$$
Eqs.(2.57)
and (2.58) are very important for any interpretation in terms of
(infinite-genus) Riemann surfaces, in particular, for identification of GKM
with appropriate Liouville-like models of $2d$ gravity. Explicit formulae for
$\cal G _W$ can be easily derived, at least, in the case of ${\cal V}={\cal
V}_K$.

\bigskip
{\it $\tau$-functions of reduced hierarchies.}
Let us note that in the fermionic language it is trivial to understand that the
$K$-reduction condition leads to the $\tau $-function $\tau ^{\{K\}}$
almost independent of times
$T_{nK}$, $i.e.$ the property (1.21). Indeed, (2.56)--(2.58) imply that the
condition of
such reduction can be transformed into the matrix  $\{{\cal G}_{mn}\}:$

$$
[{\cal G} ,E^K] = 0\hbox{  ,}
$$
where  $E$  is the shift matrix,  $(E_{mn}) = \delta _{m+1,n}$. On the same
``infinite-matrix language" the elements of  $\Gamma _\pm $ can be
parameterized as  $\gamma _+ = \exp (\sum T_kE^k)$, $\gamma _- =
\exp (\sum \tilde T_kE^{-k})$, $i.e${\it .} the currents  $J_k$ from
eq.(2.26) act as  $E^k$. Now it is evident that exponent of currents
$J_k$ with  $k = 0\ mod\ K$ can be pulled to the right vacuum and canceled,
eliminating the dependence of the corresponding times (up to the exponential
factor like that in eq.(2.54) --- this is due to the freedom to insert the
exponential  $\gamma _- = \exp (\sum \tilde T_kE^{-k})$ between  $G$
and  $\gamma _+).$

Now let us discuss the freedom in the definition of the $\tau $-function from
the viewpoint of Hirota equation. As it was pointed out in (2.54) there is a
freedom to multiply the $\tau $-function by a linear exponential of times.
Another possible modification is dealing with other components  $Gr_n$  of
Grassmannian. It corresponds to the shift of the first basis vector or,
equivalently,
multiplying it by  $\mu ^n$. All these can be trivially understood in terms of
Hirota equation in Miwa parameterization (2.42). Indeed, it is invariant with
respect to multiplying by the product  $\prod _i
f(\mu _i)$, where  $f(\mu )$  is arbitrary Laurent series. If the first term of
this series $\sim \mu ^0$, one can expand
$f(\mu _i)\rightarrow \exp \{\sum _k \mu ^{-k}_i\tilde T_k\}$  and reproduce
correct factor in eq.(2.54). If the first term behaves as $\mu ^n$, one
should deal with  $Gr_n$. Thus, in Miwa variables we readily describe all
full freedom in the definition of the $\tau $-function.

\bigskip
{\it GKM and specific points in Grassmannian.}
To conclude this subsection let us note that the problem of correspondence
between matrix models and the points of Grassmannian has been already addressed
in ref.\cite{KS91}. That paper examined the implications of  ${\cal L}_{-1}$
-constraint in the double-scaling limit of $K$-matrix model and concluded that
the corresponding point of Grassmannian is defined by the set of  $K$
linear independent vectors  $\{A^i\phi _1\}$, where operator  A  for
${{\cal V}}
= \displaystyle {{X^{K+1}\over K+1}}$
essentially coincides with ours (see (2.19), (2.20)).
Moreover, from our study of GKM we obtain straightforwardly the explicit form
of all basis vectors for {\it any} potentials: $\phi _i = \Phi _i$ as
given in eq.(2.18). Note also that in \cite{KS91} a shift of
time variables analogous to ours in (1.17) was introduced.

As to ref.\cite{KS91}, we would explain the appearance of the Airy
functions (in the case of  $K = 2$  model) from constraint (1.20) in the
following way. If we put all  $\hat T_k = 0$ except for $\hat T_1$
this constraint can be satisfied only if the
$\tau $-function is zero. This means that the relevant point of Grassmannian is
singular. Instead, we can put  $\hat T_3 = c \neq  0$  keeping all other
$\hat T_k$ with
$k\geq 5$ equal to zero. In this case we have

$$
{3\over 2} c\cdot {\partial log\ \tau \over \partial T_1} + {1\over 4} T^2_1
= 0\ .
$$
Choosing  $c = - 2/3$, so that $T_3=\hat T_3 + 2/3 = 0$, and
taking the derivative with respect to  $T_1$ and using
the relation

$$
u(T_1) =  {\partial ^2\log \ \tau \over \partial T^2_1}\ ,
$$
where  $u(T_1)$  is the KdV potential, we obtain

$$
2u(T_1) = T_1\ .
$$
According to the ideology of the inverse scattering method in the case of
2-reduction we should consider the
Schr\"odinger equation for this potential:

$$
L^2\Psi =
{\partial ^2\Psi \over \partial T^2_1} + 2u\Psi  = \mu ^2\Psi\,
$$
$i.e.$

\beq
{\partial ^2\Psi \over \partial T^2_1} + T_1\Psi  = \mu ^2\Psi\ .
\eeq
Solving this equation, we obtain Airy function

$$
\Psi (\mu |T_1)\left.\right|_{T_k\geq 3=0} = \Psi _0(\mu )
Ai (-T_1 + \mu ^2) \equiv \Psi _0(\mu ) \int e^{-x^3/3 + (\mu ^2 -T_1)x} dx
\hbox{  .}
$$
The wave function  $\Psi $  is nothing but the Baker-Akhiezer function.
$\Psi _0$ is adjusted from normalization requirment $\Psi (\mu )e^{-\sum
T_n\mu ^n} \to 1$ as $\mu \to \infty$. In
general $\Psi$ can be expressed through $\tau $-function as \cite{DJKM83}

\beq
\Psi (\mu |T_n) = e^{\Sigma T_n\mu^n} {\tau (T_n - \mu ^{-n}/n)\over \tau
(T_n)}
\eeq
(see sect.3.3 below for more details).
In eq.(2.59) we have $\Psi $-function of the form (2.60) with  $T_k = 0$, $k
\geq 3$. The admissible basis for the corresponding point of  $Gr_0$ can be
constructed by means of the $\Psi $-function as follows \cite{SW85}:

$$
\left. \phi _k(\mu ) \sim {\partial ^{k-1}\Psi \over \partial T^{k-1}_1}
\right| _{all\ T_k=0}\hbox{ .}
$$
The basis vectors, constructed according to this expression, obviously coincide
with
those which were obtained in sect.2.2 (eq.(2.6)) for the particular case of
${{\cal V}}(X) = {1\over 3} X^3.$

For generic $K$ the role of (2.59) is played by a $K$-th order differential
equations ($L^K\Psi = \mu ^2\Psi$) and the result is the Airy function of level
$K$ as defined in (2.6).

\subsection{GKM and reductions of KP-hierarchy}

We prove here that whenever potential  ${{\cal V}}(X)$  in GKM is a homogeneous
polynomial of degree  $K+1$ , $i.e$.  ${{\cal V}}(X) = const\cdot X^{K+1})$,
the partition function $Z^{\{{{\cal V}}\}}[M]$, which is a KP $\tau $-function,
in fact can be treated as a $\tau $-function  $\tau ^{\{K\}}$ of
$K$-reduced KP-hierarchy.

For this purpose let us return to the section 2.2 and note that because of
(2.7)  $F_1(\lambda )$  satisfies the Ward identity (1.23):

\beq
[{{\cal V}}'(\partial /\partial \lambda ) - \lambda ] F_1(\lambda ) = 0\hbox{.}
\eeq
If the potential  ${{\cal V}}(X)$  is a polynomial of finite degree  $K+1$ ,
then
(2.6) may be used to express  $F_{K+1}$ in the form of linear combination of
$F_i$'s  with  $i \leq  K$. Namely, if ${{\cal V}}(X) = - \sum ^{K+1}_{i=1}
{{\cal V}}_iX^i$,  $V_{K+1}= \displaystyle {{1\over K+1}}$, eq.(2.6) implies,
that

\beq
F_{K+1} = -\sum ^K_{i=1}i{{\cal V}}_i F_i + \lambda F_1.
\eeq
Clearly, all the terms on the $r.h.s$. of (2.62) except for the last one
$\lambda F_1$ will drop out of the determinant (2.9), $i.e$. in this case
$F_{K+1}$ may be defined as  $\lambda F_1$ rather than by eq.(2.62). Obviously
in the same way any  $F_{K+m}$ may be substituted by  $\lambda F_m$ , while any
$F_{nK+m} - by\ \ \lambda ^nF_m$. In other words, ${\cal F}_N[\Lambda ]$ is
given by eq.(2.9) with the first  $K$  functions  $F_1\ldots F_K$ defined
by (2.6),
while other elements of the basis are given by the recurrent relation:

\beq
F_{nK+m} \sim  \lambda ^nF_m.
\eeq
Note that this is true when  ${{\cal V}}(X)$  is {\it any} potential of degree
$K+1$, not obligatory homogeneous. However, we should recall, that  $\lambda  =
{{\cal V}}'(\mu )$  and  $\mu $  rather than  $\lambda $  is the proper
parameter
to deal with in Grassmannian picture. Therefore eq.(2.63) implies the
appearance of KP-reduction, associated with the function  ${{\cal V}}'(\mu )$.
If ${{\cal V}}$  is further restricted to be  ${{\cal V}}(X) = \displaystyle {
{X^{K+1}\over K+1}}$,
$\lambda  = \mu ^K$, and (2.63) acquires the form:

\beq
F_{nK+m} \sim  \lambda ^nF_m = \mu ^{nK}F_m,
\eeq
which is characteristic of conventional $K$-reduction.

As we already explained, $K$-reduction implies the relation (1.21) with
some factor of the form (1.22). However, even this factor is absent
in the case of GKM (1.10). For particular case of  ${{\cal V}}(X) =
const\cdot X^3$ ($K=2$) this was exhaustively proved in \cite{Kon91},
using the
properties of symplectic structure on (the model of) the universal module
space. Since interpretation of entire GKM in such terms is not yet available,
the analog of such proof for generic $K$ is still lacking.

\subsection{On $T_{nK}$-independence of $Z^{\{K\}}$}

{\it Reformulation of the problem.} First of all, let us choose our variables
$\{\mu _i\}$  in such a way that the only non-vanishing times  $T_n= {1\over n}
\sum _i \mu ^{-n}_i$ are those with  $n = 0\ mod\ K$. To do this it is
enough to have only  $K$  finite parameters  $\mu _1,\ldots,\mu _K$, which are
essentially  $K$-th order roots of unity:

\beq
\mu _j = \mu \varepsilon ^j\hbox{, } \varepsilon= e^{2\pi i/K}\hbox{, }   j =
1\ldots K\hbox{.}
\eeq
For such choice of variables the formula (1.21),

\bea
\tau ^{\{K\}}(T_1\ldots T_{K-1}\hbox{, } T_K\hbox{, } T_{K+1}\ldots \hbox{. }
T_{2K-1}\hbox{, } T_{2K}\hbox{, } T_{2K+1}\ldots ) =\nn \\
= e ^{\sum _na_{nK}T_{nK}} \tau ^{\{K\}}(T_1\ldots T_{K-1}\hbox{, 0, }
T_{K+1}\ldots \hbox{. }
T_{2K-1}\hbox{, 0, } T_{2K+1}\ldots ),
\eea
turns into:

\beq
\tau ^{\{K\}}(0\ldots 0,T_K,0\ldots 0,T_{2K},0\ldots )
= e ^{\sum _na_{nK}T_{nK}} \tau ^{\{K\}}(0) = e ^{ \sum _n
{a_{nK} \over n}
\mu ^{-nK}}\tau ^{\{K\}}(0).
\eeq
This relation is valid for {\it any} $\tau $-function  $\tau ^{\{K\}}$ of
$K$-reduced
KP hierarchy. Our purpose is to prove that for the {\it particular}
example of such
$\tau ^{\{K\}}$ , namely, for partition function $Z^{\{K\}}$ of GKM, associated
with potential of the form  ${{\cal V}}(X) = X^{K+1}/(K+1)$ , all the

\beq
a_{nK}[Z^{\{K\}}] = \partial \ \log \ Z^{\{K\}}/\partial T_{nK} = 0,
\eeq
or, equivalently, the function in the $r.h.s$. of (2.67) is independent of
$\mu $. In other words, we need to prove that

\beq
\xi ^{\{K\}}(\mu ) \equiv  Z^{\{K\}}[\mu _j]\left.\right|_{(2.65)} = 1.
\eeq
The $l.h.s$. of this relation can be evaluated with the help of eq.(2.16),

$$
Z^{\{K\}} = {{\rm det} _{(ij)}\Phi _i(\mu _j)\over \Delta (M)}.
$$
We emphasize that in order to prove  $T_{nK}$- independence of  $Z^{\{K\}}$ it
is enough to examine the determinant of  $K\times K$  matrix in (2.16) with
entries

\beq
\Phi _i(\mu ) =
\left( s(\lambda ){\partial \over \partial \lambda }s(\lambda )^{-1}\right) ^{i
-1} \Phi _1(\mu ) \equiv  A^{i-1}_{\{K\}} \Phi _1 =
\mu ^{i-1}e^{(i)}(\mu ),
\eeq
\beq
e^{(i)}(\mu ) =\sum _{\alpha \geq 0}
e^{(i)}_\alpha \mu ^{-\alpha (K+1)}\hbox{, }   e^{(i)}_0 = 1
\eeq
(of course, all the quantities  $\Phi $, $s$, $A$  (see (2.15),
(2.18)-(2.20)) and  $e^{(i)}$ depend on the form of potential  ${{\cal V}}$
and
thus on  $K).$

\bigskip
{\it Straightforward calculation $(K = 2)$.} We begin our proof that
$\xi ^{\{K\}}(\mu ) = 1$  from direct calculation of this quantity. If we
substitute the expansion (2.71) with  $\mu$ 's  defined in (2.65) into (2.16)
the result reads:

\beq
\xi ^{\{K\}}(\mu ) =\sum _{\{\alpha _i\}}\left( \prod _{i=1}^K
e^{(i)}_{\alpha _i}\right) \cdot {1\over \mu ^{\Sigma
\alpha _i}}\cdot {{\rm det} _{
(ij)}\varepsilon ^{(-\alpha _i+i-1)j}\over {\rm det} _{(ij)}
\varepsilon ^{(i-1)j}}.
\eeq
The remaining determinant at the $r.h.s$. is actually vanishing unless
$\sum _i \alpha _i = 0\ mod\ K$, so that (2.68) is in fact equivalent to the
set of identities

\beq
\sum _{{\{\alpha _i\}}\atop {\sum \{\alpha _i\}=nK}}\left( \prod _{i=1}^K
e^{(i)}_{\alpha _i}\right) \cdot {{\rm det} _{(ij)}\varepsilon
^{(-\alpha _i+i-1)j}
\over {\rm det} _{(ij)}\varepsilon ^{(i-1)j}}= \delta _{n,0}\hbox{ for all }\ n
\geq  0
\eeq
for coefficients  $e^{(i)}_\alpha $ of series expansions of modified $K$-level
Airy functions

\beq
\Phi ^{\{K\}}_i(\mu ) = \mu ^{{K-1\over 2}} e^{-
{K\mu ^{K+1}\over K+1}}\int e^{- {x^{K+1}\over K+1}+ \mu ^Kx}x^{i-1}dx.
\eeq
For example, for $K = 2$ identities (2.73) look like

\beq
\sum _{\alpha +\beta =2n}(-)^\alpha e^{(1)}_\alpha e^{(2)}_\beta  =
\delta _{n,0}.
\eeq
Since for $K = 2$

\bea
\Phi _1(\mu ) & = & \sqrt{{\mu \over \pi }} e^{- {2\over 3}\mu ^3}\int
dxe^{- {x^3\over 3} + x\mu ^2}
= {1\over \sqrt{\pi }} \int   dz\ \exp \left\lbrace - z^2-
{z^3\over 3\mu ^{3/2}}\right\rbrace  =\nn \\
& = & \sum  _k {1\over 9^k(2k)!}
{\Gamma (3k+1/2)\over \Gamma (1/2)} {1\over \mu ^{3k}}\hbox{  ,}\nn \\
\Phi _2(\mu ) & = & \sqrt{{\mu \over \pi }} e^{- {2\over 3}\mu ^3}\int
dx\cdot x\ e^{- {x^3\over 3} + x\mu ^2} =
{1\over \sqrt{\pi }} \int   dz (z + \mu )\exp \left\lbrace - z^2 -
{z^3\over 3\mu ^{3/2}}\right\rbrace =\nn \\ & = & - \sum  _k {1\over 9^k(2k)!}
{6k+1\over 6k-1} {\Gamma (3k+1/2)\over \Gamma (1/2)} {1\over \mu ^{3k}}\ ,\nn
\eea
$i.e$

$$
e^{(1)}_\alpha  = {1\over 9^\alpha (2\alpha )!} {\Gamma (3\alpha +1/2)\over
\Gamma (1/2)}\hbox{
, }  e^{(2)}_\alpha  = - {6\alpha +1\over 6\alpha -1} e^{(1)}_\alpha\ .
$$
A more explicit form of (2.75) is

\beq
\sum_{{\alpha +\beta =2n}\atop {\alpha ,\beta \geq 0}}
{36\alpha \beta  - 1\over (2\alpha )!(2\beta )!}
\Gamma (3\alpha -1/2) \Gamma (3\beta -1/2) = - 4[\Gamma (1/2)]^2\delta _{n,0}\
,
\eeq
which is indeed a valid $\Gamma$-function identity.

This calculation, though absolutely straightforward, is hard to repeat for
generic $K$. Still it is useful for illustrative purposes and we included it
into this section.

\bigskip
{\it The proof.} A much easier approach is to prove that

\beq
\mu  {\partial \over \partial \mu } \xi ^{\{K\}}(\mu ) = 0.
\eeq
Since it is obvious from (2.72) that at least  $\xi ^{\{K\}} \rightarrow  1$
as  $\mu  \rightarrow  \infty $, this would provide a complete proof of (2.69).
In order to prove (2.78) it is enough to act with
$\mu \partial /\partial \mu $  upon

\beq
\xi ^{\{K\}}(\mu ) \sim  {\rm det} _{(ij)}\varepsilon ^{ij}e^{(i)}(\mu
\varepsilon
^j)
\eeq
and then make use of the fact, that  $\mu {\partial \over \partial \mu }
e^{(i)}(\mu \varepsilon ^j)$  can be decomposed into linear combination of
$e^{(1)}(\mu \varepsilon ^j)\ldots e^{(K)}(\mu \varepsilon ^j)$.
This decomposition, which
is, of course, the implication of $K$-reduction, can be worked out from the
relation (2.20),

\beq
\mu ^ie^{(i+1)}(\mu ) = A_{\{K\}}\ \mu ^{i-1}e^{(i)}(\mu ).
\eeq
Since

\beq
A_{\{K\}} = {1\over K\mu ^{K-1}} {\partial \over \partial \mu } + \mu  -
{K-1\over 2K\mu ^K},
\eeq
(see also \cite{KS91}), we have:

\beq
(\mu {\partial \over \partial \mu })e^{(i)}(\mu \varepsilon ^j) = ({K+1-2i
\over 2}
- K\varepsilon ^j\mu ^{K+1})e^{(i)}(\mu \varepsilon ^j)  +
K\varepsilon ^j\mu ^{K+1}e^{(i+1)}(\mu \varepsilon ^j).
\eeq
This should be supplemented by the condition

\beq
e^{(K+1)} = e^{(1)}
\eeq
(equation  $\lambda e^{(1)} = \mu ^Ke^{(1)} = A_{\{K\}}^Ke^{(1)} =
s(\lambda )\left(\displaystyle {
 {\partial \over \partial \lambda }}\right) ^Ks(\lambda )^{-1}e
^{(1)}$ is nothing but the equation for the level-$K$ Airy function
$F^{\{K\}}_1 = s^{-1}e^{(1)} = \int   e^{- {x^{K+1}\over K+1}+ \mu ^Kx}dx).$

Let us begin our study of (2.78) from the familiar example of  $K = 2$. First
of all, one can check up (2.82) with the help of explicit expressions (2.76)
for  $e^{(i)}_\alpha $. Then

$$
(\mu {\partial \over \partial \mu }) \xi ^{\{2\}}(\mu ) \sim
(\mu {\partial \over \partial \mu })(e^{(1)}(\mu )e^{(2)}(-\mu ) +
e^{(2)}(\mu )e^{(1)}(-\mu )) =
$$
$$
= [({1\over 2} -2\mu ^3)e^{(1)}(\mu ) + 2\mu ^3e^{(2)}(\mu )]e^{(2)}(-\mu ) +
e^{(1)}(\mu )[-2\mu ^3e^{(1)}(-\mu ) + (-{1\over 2} + 2\mu ^3)e^{(2)}(-\mu )]
+
$$
$$
+ [2\mu ^3e^{(1)}(\mu ) + (-{1\over 2} - 2\mu ^3)e^{(2)}(\mu )]e^{(1)}(-\mu ) +
e^{(2)}(\mu )[({1\over 2} + 2\mu ^3)e^{(1)}(-\mu ) - 2\mu ^3e^{(2)}(-\mu )],
$$
and the $r.h.s$. is identically zero.

If we return to the case of arbitrary  $K$ , note that after the substitution
of (2.79) into the $l.h.s$. of (2.78) we obtain a sum of  $K$  determinants
like (2.79) with $\varepsilon ^{i'j}e^{(i')}(\mu \varepsilon ^j)$
in the row  $i '$
substituted by (2.82):

$$
(\mu {\partial \over \partial \mu })\varepsilon ^{i'j}e^{(i')}(\mu
\varepsilon ^j) =
({K+1-2i'\over 2} - K\varepsilon ^j\mu ^{K+1})\varepsilon ^{i'j}e^{(i')}(\mu
\varepsilon ^j)
+  K\mu ^{K+1}\varepsilon ^{(i'+1)j}e^{(i'+1)}(\mu \varepsilon ^j).
$$
Next, note that the last item in the $r.h.s$. coincides with the entry
$\varepsilon ^{(i'+1)j}e^{(i'+1)}(\mu \varepsilon ^j)$ of the next line of the
{\it same}
determinant, and thus can be eliminated. Moreover, the first item at the
$r.h.s$. implies just that the  $i'$-th line is multiplied by a factor of
$\displaystyle {{K+1-2i'\over 2}}$
and the effect of this cancels in the sum over  $i'$
: $\sum _{i'=1}^K\displaystyle {{K+1-2i'\over 2}} = 0$.
Therefore, we conclude that

\beq
\mu {\partial \over \partial \mu } \xi ^{\{K\}}(\mu ) \sim
\mu {\partial \over \partial \mu }
{\rm det} _{(ij)}\varepsilon ^{ij}e^{(i)}(\mu \varepsilon ^j) \equiv
\mu {\partial \over \partial \mu } {\cal D} =
-K\mu ^{K+1}\sum ^K_{i'=1}{\cal D}_{i'}\ ,
\eeq
where ${\cal D}_{i'}$ is just the same determinant  ${\cal D}$ , only in the
$i'$-th line  $d_{i'j} \equiv \varepsilon ^{i'j}e^{(i')}(\mu \varepsilon ^j)$
is
substituted by  $\varepsilon ^jd_{i'j} = \varepsilon ^{(i'+1)j}e^{(i')}(\mu
\varepsilon ^j)$.
The sum of such  ${\cal D}_{i'}$ (for {\it any} original ${\cal D} )$ is
identical zero as a consequence of the identity  $\sum ^K_{j=1}\varepsilon ^j
= 0.$

\bigskip
{\it Examples:}

$K = 2:$

$$
{\cal D} = {\rm det}  \left( \begin{array}{cc}d_{11} & d_{12}\\ d_{21} & d_{22}
\\
\end{array}\right)  =
d_{11}d_{22} - d_{12}d_{21};
$$

$$
{\cal D}_1 = {\rm det}  \left( \begin{array}{cc}-d_{11} & (-)^2d_{12}\\
d_{21} & d_{22}\\ \end{array}\right)
= - d_{11}d_{22} - d_{12}d_{21}\hbox{, }
$$

$$
{\cal D}_2 = {\rm det}
\left( \begin{array}{cc}d_{11} & d_{12}\\ -d_{21} & (-)^2d_{22}\\ \end{array}
\right)  = d_{11}d_{22} + d_{12}d_{21},
$$
and

$$
{\cal D}_1 + {\cal D}_2 = 0.
$$
\bigskip

$K = 3:  \varepsilon = e^{2\pi i/3}$,

$$
{\cal D} =  {\rm det}  \left( \begin{array}{ccc}d_{11} & d_{12} & d_{13}\\
d_{21} & d_{22} & d_{23}\\ d_{31} & d_{32} & d_{33}\\ \end{array}
\right)   =
$$

$$
= d_{11}d_{22}d_{33} + d_{12}d_{23}d_{31} + d_{13}d_{32}d_{21} -
d_{12}d_{21}d_{33} - d_{13}d_{31}d_{22} - d_{23}d_{32}d_{11},
$$

$$
{\cal D}_1=  {\rm det}  \left( \begin{array}{ccc}
\varepsilon d_{11} & \varepsilon ^2d_{12} & \varepsilon ^3d_{13}\\
d_{21} & d_{22} & d_{23}\\ d_{31} & d_{32} & d_{33}\\ \end{array}
\right)  =
$$

$$
= \varepsilon d_{11}d_{22}d_{33} + \varepsilon ^2d_{12}d_{23}d_{31} +
\varepsilon ^3d_{13}d_{32}d_{21} - \varepsilon ^2d_{12}d_{21}d_{33} -
\varepsilon ^3d_{13}d_{31}d_{22} - \varepsilon d_{23}d_{32}d_{11},
$$

$$
{\cal D}_2 =  {\rm det}  \left(
\begin{array}{ccc}d_{11} & d_{12} & d_{13}\\
\varepsilon d_{21} & \varepsilon ^2d_{22} & \varepsilon ^3d_{23}\\
d_{31} & d_{32} & d_{33}\\ \end{array}
\right)  =
$$

$$
= \varepsilon ^2d_{11}d_{22}d_{33} + \varepsilon ^3d_{12}d_{23}d_{31} +
\varepsilon d_{13}d_{32}d_{21} - \varepsilon d_{12}d_{21}d_{33} -
\varepsilon ^2d_{13}d_{31}d_{22} - \varepsilon ^3d_{23}d_{32}d_{11},
$$

$$
{\cal D}_3 =  {\rm det}  \left(
\begin{array}{ccc}d_{11} & d_{12} & d_{13}\\
d_{21} & d_{22} & d_{23}\\
\varepsilon d_{31} & \varepsilon 2d_{32} & \varepsilon ^3d_{33}\\ \end{array}
\right)  =
$$

$$
= \varepsilon ^3d_{11}d_{22}d_{33} + \varepsilon d_{12}d_{23}d_{31} +
\varepsilon ^2d_{13}d_{32}d_{21} - \varepsilon ^3d_{12}d_{21}d_{33} -
\varepsilon d_{13}d_{31}d_{22} - \varepsilon ^2d_{23}d_{32}d_{11},
$$
and

$$
{\cal D}_1 + {\cal D}_2 + {\cal D}_3 = 0.
$$

\bigskip
These two examples are enough to illustrate the general phenomenon. The thing
is that determinant  ${\cal D}$  is an algebraic sum of products  $d_{i_1j_2}
\ldots d_{i_Kj_K}$. Let us pick up one of these items. It appears in
${\cal D}_{i'}$ with the coefficient  $\varepsilon ^{j'}$ , where  $j'$  is
defined
as the second subscript of the element  $d_{i'j'}$ in this product. The map
$i'\rightarrow  j'$ depends on the particular item we choose, but when we sum
over all  $i'$  this is the same as to sum over all  $j'$ , since every
subscript appears in the product once and only once. Therefore in the sum
$\sum _{i'}{\cal D}_{i'}$ every item appears with the universal coefficient
$\sum _{j'}\varepsilon ^{j'} = 0$ , $i.e$. $\sum _{i'}{\cal D} _{i'} = 0.$

This completes our proof of (2.78) and thus of (2.68).

\bigskip
{\it Remark.} To avoid possible confusion, let us note that it is important
that Airy-functions are analytic functions of  $\mu $ , but {\it not} of
$\lambda  = \mu ^K (e.g.$,  $e^{(i)}(\mu )$  are expanded in series with
integer
powers of  $\mu ^{-(K+1)} = \lambda ^{-1-1/K})$. This is what makes the proof
slightly non-trivial. Otherwise, if one suggested that the entries of the
matrix in (2.79) are analytic functions of  $\lambda $ , the
$\lambda $-derivative of this determinant would be simply a combination of
determinants with coincident rows and thus vanishing. However, non-analiticity
in  $\lambda $  requires one to be more accurate:  $\lambda ^{1/K}$ takes
different values  $\mu \varepsilon ^j$ when appearing in different places, and
a detailed calculation, as given above, is necessary.

\section{Universal ${\cal L}_{-1}$-constraint and string equation}
\setcounter{equation}{0}
\subsection{Motivations}

Since it was proved in sect.2 that  $Z^{\{K\}}[M]$  is a $\tau $-function
$\tau ^{\{K\}}$ of the  $K$-reduced KP hierarchy, in order to define it
completely $(i.e$. to fix the point of Grassmannian) it is enough to deduce
somehow a single additional constraint

\beq
{\cal L}^{\{K\}}_{-1}Z^{\{K\}} = 0,
\eeq
with

\beq
{\cal L}^{\{K\}}_{-1} = {\cal W}^{(2)}_{-K} = {1\over K}
\sum _{{n\geq 1}\atop {n \ne 0\ mod\ K}}
(n+ K)T_{n+K}\partial /\partial T_n - \partial /\partial T_1 + {1\over 2K}
\sum _{{a+b=K}\atop {a,b \geq 1}}aT_abT_b.
\eeq
According to arguments of ref.\cite{FKN91} such constraint (3.1),
when imposed on
$Z^{\{K\}} = \tau ^{\{K\}}$, implies the entire set of  $W_K$-algebra
constraints in the form of (1.29). Another reason for the study of
${\cal L}_{-1}$- constraint is that it is much simpler than any other ones:
${\cal L}_{-1}$ is the only operator of interest which does not contain double-
and higher- order $T$-derivatives (to be exact, there is one more such
generator:  ${\cal L}_0 )$. Among other things, this means that it is easier to
formulate the   {\it universal   $(i.e$}. for {\it any} potential
${{\cal V}}(X)$) ${\cal L}_{-1}$-constraint, than any other corollary of
(universal!) Ward-identity (1.23).

This section contains formulation and derivation of
${\cal L}_{-1}^{\{{\cal V}\}}$-constraint
for GKM with arbitrary potential  ${{\cal V}}(X)$ , making use of
explicit formulas, derived in sect.2. Namely, we are going to exploit the fact
that the functions
$\tilde \Phi _i(\mu )$  in (2.10) are not independent, but are
rather related by the action of  $\partial /\partial \lambda $- operator: see
(2.20). The proof itself is described in sect.3.2 and sect.3.3 contains
additional remarks about string equation. We emphasize, that not only the
${\cal L}_{-1}$- constraint is valid for any  ${{\cal V}}(X)$  (with the only
restriction that  ${{\cal V}}(x)$  grows faster than  $x$  as  $x \rightarrow
\infty )$, it is just the same for any size  $N$  of the matrices. Just like
the property of integrability $(i.e$. that  $Z^{\{{{\cal V}}\}}=
\tau ^{\{{{\cal V}}\}})$, this constraint is not sensitive to  $N$ , and this
makes the entire construction behaving smoothly in continuum limit, as  $N
\rightarrow  \infty .$

\subsection{Direct evaluation of ${\cal L}_{-1}Z$}

It is well known \cite{Orl88,GMMMO91}, that  ${\cal L}_{-1}$-constraint is
closely
related to the action of operator

\beq
Tr{\partial \over \partial \Lambda _{tr}} = Tr {1\over {{\cal V}}''(M)}
{\partial \over \partial M_t}_r.
\eeq
Therefore it is natural to examine, how this operator acts on

\beq
Z^{\{{{\cal V}}\}}[M] = {{\rm det} \
\tilde \Phi _i(\mu _j)\over \Delta (M)}\prod _i
s(\mu _i),
\eeq

\beq
s(\mu ) = ({{\cal V}}''(\mu ))^{1/2}e^{{{\cal V}}(\mu )-\mu {{\cal V}}'(\mu )},
\eeq

$$
\tilde \Phi _i(\mu ) = F_i(\lambda ) =
(\partial /\partial \lambda )^{i-1}F_1(\lambda )\hbox{, }   \lambda  =
{{\cal V}}'(\mu ).
$$
First of all, if $Z^{\{{{\cal V}}\}}$ is considered as a function
of  $T$-variables,

\beq
{1\over Z^{\{{{\cal V}}\}}}Tr{\partial \over \partial \Lambda _{tr}}
Z^{\{{{\cal V}}\}} = -\sum _{n\geq 1}Tr [{1\over {{\cal V}}''(M)M^{n+1}}]
{\partial logZ^{\{{{\cal V}}\}}\over \partial T_n}.
\eeq
In the particular case of  ${{\cal V}}(X) = {{\cal V}}_K(X) = \displaystyle {
{X^{K+1}\over K+1}}$ ,
eq.(3.6) turns into

\bea
{1\over Z^{\{K\}}} Tr{\partial \over \partial \Lambda _{tr}}Z^{\{K\}} =
 - {1\over KZ^{\{K\}}}
\sum _{{n\geq 1}\atop {n \ne 0\ mod\ K}}
Tr{1\over M^{n+K}} {\partial Z^{\{K\}}\over \partial T_n}= \nn \\
= - {1\over Z^{\{K\}}} \left\lbrace {\cal L}^{\{K\}}_{-1} - {1\over 2}
\sum _{{a+b=K}\atop {a,b>1}}aT_abT_b +
{\partial \over \partial T_1}\right\rbrace  Z^{\{K\}}.
\eea
We made use of the definition (3.2) of  ${\cal L}^{\{K\}}_{-1}$ -operator and
the fact that  $Z^{\{K\}}$ is independent of all  $T_{nK}$ (since it is a
$\tau ^{\{K\}}$ -function). Time-variables are defined by Miwa transformation
$T_n = {1\over n} Tr\ M^{-n}.$
On the other hand, if we apply (3.3) to explicit formula (3.4), we obtain:

\beq
\new
\begin{array}{c}
{1\over Z^{\{{{\cal V}}\}}}Tr{\partial \over \partial \Lambda _{tr}}
Z^{\{{{\cal V}}\}}\\
 = - Tr\ M + {1\over 2} \sum _{i,j}{1\over
{{\cal V}}''(\mu _i){{\cal V}}''(\mu _j)}
{{{\cal V}}''(\mu _i)-{{\cal V}}''(\mu _j)\over \mu _i - \mu _j} +
Tr{\partial \over \partial \Lambda _{tr}}\log \ {\rm det} \ F_i(\lambda _j),
\end{array}
\eeq
or, in the case of  ${{\cal V}}(X) = X^{K+1}/(K+1)$ ,

\beq
{1\over Z^{\{K\}}}Tr{\partial \over \partial \Lambda _{tr}}Z^{\{K\}} = - TrM +
{1\over 2}
\sum _{{a+b=K}\atop {a,b>1}}aT_abT_b +
Tr{\partial \over \partial \Lambda _{tr}}\log \ {\rm det} \ F_i(\lambda _j).
\eeq
Combination of (3.7) and (3.9) implies, that

\beq
{1\over Z^{\{K\}}}{\cal L}^{\{K\}}_{-1}Z^{\{K\}} = -
{\partial \over \partial T_1}\log \ Z^{\{K\}} + TrM -
Tr{\partial \over \partial \Lambda _{tr}}\log \ {\rm det} \ F_i(\lambda _j).
\eeq
The $r.h.s$. here is practically independent of  $K$ , and this may be used,
together with eqs.(3.6) and (3.8) in order to suggest the formula for the
universal operator  ${\cal L}^{\{{{\cal V}}\}}_{-1} :$ the idea is to preserve
the relation (3.10):

\beq
{1\over Z^{\{{{\cal V}}\}}}{\cal L}^{\{{{\cal V}}\}}_{-1} Z^{\{{{\cal V}}\}}_N
= - {\partial \over \partial T_1}\log \ Z^{\{{{\cal V}}\}}_N + TrM -
Tr{\partial \over \partial \Lambda _{tr}}\log \ {\rm det} \ F_i(\lambda _j).
\eeq
Here

\bea
{\cal L}^{\{{{\cal V}} \}}_{-1} =\sum _{n\geq 1}Tr
[{1\over {{\cal V}}''(M)M^{n+1}}] {\partial \over \partial T_n} + \nn \\
+ {1\over 2}
\sum _{i,j}{1\over {{\cal V}}''(\mu _i){{\cal V}}''(\mu _j)}{{{\cal V}}''
(\mu _i)-
{{\cal V}}''(\mu _j)\over \mu _i - \mu _j} - {\partial \over \partial T_1},
\eea
and this turns into (3.2) when  ${{\cal V}}(X) = X^{K+1}/(K+1)$  (note that the
items with  $i=j$  are included into the sum at the $r.h.s$. in (3.12)).

So, in order to prove the ${\cal L}^{\{{{\cal V}} \}}_{-1}$-constraint, one
should prove that the $r.h.s$. of (3.9) vanishes,

\beq
{\partial \over \partial T_1}\log \ Z^{\{{{\cal V}}\}}_N = TrM -
Tr{\partial \over \partial \Lambda _{tr}}\log \ {\rm det} \ F_i(\lambda _j)
\hbox{,}
\eeq
for  $Z^{\{{{\cal V}}\}}_N = \tau ^{\{{{\cal V}}\}}_N$, defined by (3.4). The
$l.h.s$. may be represented as residue of the ratio

\beq
res_\mu {\tau ^{\{{{\cal V}}\}}_N(T_n+
\mu ^{-n}/n)\over \tau ^{\{{{\cal V}}\}}_N(T_n)} =
{\partial \over \partial T_1}\log \ \tau ^{\{{{\cal V}}\}}_N(T_n).
\eeq
However, if expressed through Miwa coordinates, the $\tau $-function in the
numerator is given by the same formula (3.4) with one extra parameter  $\mu $ ,
$i.e$. is in fact equal to $\tau ^{\{{{\cal V}}\}}_{N+1}$ . This idea is almost
enough to deduce (3.13). Since (3.13) is valid for any value of $N$, it is
reasonable to begin with an illustrative example of  $N = 1$. Then $(\lambda  =
{{\cal V}}'(\mu ))$

$$
\tau ^{\{{{\cal V}}\}}_1(T_n) = \tau ^{\{{{\cal V}}\}}_1[\mu _1] =
e^{{{\cal V}}(\mu _1)-\mu _1{{\cal V}}'(\mu _1)}[{{\cal V}''}(\mu _1)]^{1/2}F(
\lambda _1){ , }
$$

\beq
\new
\begin{array}{c}
\tau ^{\{{{\cal V}}\}}_1(T_n+\mu ^{-n}/n) =
\tau ^{\{{{\cal V}}\}}_2[\mu _1,\mu ] = \\
= e^{{{\cal V}}(\mu _1)-\mu _1{{\cal V}}'(\mu _1)}e^{{{\cal V}}(\mu )-\mu
{{\cal V}}'(
\mu )} {[{{\cal V''}}(\mu _1){{\cal V''}}(\mu )]^{1/2}\over \mu  -
\mu _1}[F(\lambda _1)\partial F(\lambda )/\partial \lambda  -
F(\lambda )\partial F(\lambda _1)/\partial \lambda _1] = \\
= {e^{{{\cal V}}(\mu )-\mu {{\cal V}}'(\mu )}[{{\cal V''}}(\mu )]^{1/2}
F(\lambda )\over \mu  - \mu _1} \tau ^{\{{{\cal V}}\}}_1[\mu _1]\cdot [ -
\partial logF(\lambda _1)/\partial \lambda _1 +
\partial logF(\lambda )/\partial \lambda ].
\end{array}
\eeq
The function

\beq
F(\lambda ) =\int   dx\ e^{-{{\cal V}}(x)+\lambda x} \sim
e^{{{\cal V}}(\mu )-\mu {{\cal V}}'(\mu )} [{{\cal V}}''(\mu )]^{-1/2}\{1 +
O({{{\cal V}}''''\over {{\cal V}}''{{\cal V}}''})\}.
\eeq
If  ${\cal V}(\mu )$  grows with  $\mu ^n$ as  $\mu  \rightarrow  \infty $ ,
then
${\cal V}''''/({\cal V}'')^2 \sim  \mu ^{-n}$ , and for our purposes it is
enough to have  $n > 1$ , so that in the braces at the $r.h.s$. stands
$\{1+o(1/\mu )\}  (\mu \cdot o(\mu ) \rightarrow  0$ as $\mu
\rightarrow  \infty )$. Then numerator at the $r.h.s$. of (3.15) is $\sim  1 +
o(1/\mu )$, while the second item in square brackets behaves as
$\partial logF(\lambda )/\partial \lambda  \sim  \mu (1+o(1/\mu ))$.
Combining all this with eq.(3.14), we obtain:

\bea
{\partial \over \partial T_1}\log \ \tau ^{\{{\cal V}\}}_1 & = &
res_\mu \left\lbrace  {1+o(1/\mu )\over \mu  - \mu _1} [-
\partial logF(\lambda _1)/\partial \lambda _1 +
\mu (1+o(1/\mu ))]\right\rbrace  = \nn \\
& = & \mu _1 - \partial logF(\lambda _1)/\partial \lambda _1.
\eea
Thus (3.13) is proved for the particular case of  $N =1.$

The proof is literally the same for any  $N$ , only instead of a relatively
simple expression in square brackets at the $r.h.s$. of (3.15) one has the
ratio:

\beq
{\rm det}  \left|  \begin{array}{ccccc}
F(\lambda _1) & \partial F(\lambda _1)/\partial \lambda _1 & \ldots
& \partial ^{N-1}F(\lambda _1)/\partial \lambda _1 ^{N-1}
& \partial ^NF(\lambda _1)/\partial \lambda _1 ^N\\
F(\lambda _2) & \partial F(\lambda _2)/\partial \lambda _2 & \ldots
& \partial ^{N-1}F(\lambda _2)/\partial \lambda _2 ^{N-1}
& \partial ^NF(\lambda _2)/\partial \lambda _2 ^N\\
\vdots & \vdots & \ddots & \vdots & \vdots\\
F(\lambda _N) & \partial F(\lambda _N)/\partial \lambda _N & \ldots
& \partial ^{N-1}F(\lambda _N)/\partial \lambda _N ^{N-1}
& \partial ^NF(\lambda _N)/\partial \lambda _N ^N\\
F(\lambda ) & \partial F(\lambda )/\partial \lambda  & \ldots
& \partial ^{N-1}F(\lambda )/\partial \lambda  ^{N-1}
& \partial ^NF(\lambda )/\partial \lambda  ^N\\ \end{array}
 \right|
\eeq
over

\beq
{\rm det}  \left|  \begin{array}{cccc}
F(\lambda _1) & \partial F(\lambda _1)/\partial \lambda _1 & \ldots
& \partial ^{N-1}F(\lambda _1)/\partial \lambda _1 ^{N-1} \\
F(\lambda _2) & \partial F(\lambda _2)/\partial \lambda _2 & \ldots
& \partial ^{N-1}F(\lambda _2)/\partial \lambda _2 ^{N-1}  \\
\vdots & \vdots & \ddots & \vdots \\
F(\lambda _N) & \partial F(\lambda _N)/\partial \lambda _N & \ldots
& \partial ^{N-1}F(\lambda _N)/\partial \lambda _N ^{N-1}
\\ \end{array}
 \right|\ ,
\eeq
which is in fact equal to

\beq
F(\lambda )\mu ^N\left\lbrace [1+o(1/\mu )] - {1\over \mu }
[Tr{\partial \over \partial \Lambda _{tr}}\log \ {\rm det} \ F_i(\lambda _j)]
\cdot [
1+{\cal O}(1/\mu )]\right\rbrace \hbox{.}
\eeq
We used here the estimates  $\displaystyle {{\partial F/\partial
\lambda \over F}} =
\mu (1+o(1/\mu ))$ ,  $\displaystyle {{\partial ^2F/\partial \lambda ^2\over F}
=
\left\lbrace {\partial F/\partial \lambda \over F}\right\rbrace ^2+
{\partial \over \partial \lambda }}[\mu (1+o(1/\mu ))] =
\mu ^2(1+o(1/\mu ))$, ... , $\displaystyle {{\partial ^NF/\partial
\lambda ^N\over F}}=
\mu ^N(1+o(1/\mu ))$, which allow us to pick up only the contributions
with  $\partial ^NF(\lambda )/\partial \lambda ^N$ and
$\partial ^{N-1}F(\lambda )/\partial \lambda ^{N-1}$ to (3.18) --- all other
are lower order in  $\mu $  as  $\mu  \rightarrow  \infty $. The  $N$-th
derivative in (3.18) is multiplied by determinant of  $N\times N$  matrix,
which is exactly (3.19), while the $(N-1)$-th derivative --- by

\beq
{\rm det}  \left|  \begin{array}{ccccc}
F(\lambda _1) & \partial F(\lambda _1)/\partial \lambda _1 & \ldots
& \partial ^{N-2}F(\lambda _1)/\partial \lambda _1 ^{N-2}
& \partial ^NF(\lambda _1)/\partial \lambda _1 ^N\\
F(\lambda _2) & \partial F(\lambda _2)/\partial \lambda _2 & \ldots
& \partial ^{N-2}F(\lambda _2)/\partial \lambda _2 ^{N-2}
& \partial ^NF(\lambda _2)/\partial \lambda _2 ^N\\
\vdots & \vdots & \ddots & \vdots & \vdots\\
F(\lambda _N) & \partial F(\lambda _N)/\partial \lambda _N & \ldots
& \partial ^{N-2}F(\lambda _N)/\partial \lambda _N ^{N-2}
& \partial ^NF(\lambda _N)/\partial \lambda _N ^N
\\ \end{array}
 \right|
\eeq
This is, however, nothing but  $Tr\displaystyle {
{\partial \over \partial \Lambda _{tr}}}$  of
the logarithm of (3.19), and for this to be true it is absolutely essential,
that  $F_i\sim  (\partial /\partial \lambda )^{i-1}F_1.$

With the estimate (3.20) we obtain from (3.14):

\bea
\lefteqn{{\partial \over \partial T_1}\log \ \tau ^{\{{\cal V}\}}_N = }\nn \\
& & = res_\mu \left\lbrace {1+o(1/\mu )\over  \prod _{j=1}
(\mu -\mu _j)} \mu ^N\left\lbrace [1+o(1/\mu )] \right. \right. - \nn \\ & & -
\left. \left.
{1\over \mu }[Tr{\partial \over \partial \Lambda _{tr}}\log \ {\rm det} \ F_i(
\lambda _j)]\cdot [1+{\cal O}(1/\mu )]\right\rbrace \right\rbrace = \nn \\
& & =\sum ^N_{j=1}\mu _j -
Tr{\partial \over \partial \Lambda _{tr}}\log \ {\rm det} \ F_i(\lambda _j).
\eea
This completes the proof of eq.(3.13) and thus of the universal
${\cal L}^{\{{\cal V}\}}_{-1}$-constraint:

\beq
{\cal L}^{\{{\cal V}\}}_{-1} \tau ^{\{{\cal V}\}} = 0.
\eeq

We proved the crucial
eq.(3.13) by the direct calculation. Now we would like to
describe an alternative proof which originates from representation of
$\tau $-function through the current correlators described in sect.2.3. This
approach may be useful for evaluation of the higher $T$-derivatives
 (see sect.3.4 below). Namely,
from eqs.(2.35) and (2.38) it is obvious that  $\tau ^G_N[t|M] \equiv
\tau (t,\{T_n\})$  depends on  $t$  and  $T_n$ as follows:
$\tau (t,\{T_n\}) = \tau (T_1 - Nt\hbox{, } T_2 - {1\over 2}Nt^2\hbox{, ...)}$
$i.e$. in eq.(2.38)

\beq
\tau ^G_N[t|M] = \exp (-N\sum {t^n\over n}{\partial \over \partial T_n})
\tau (T) = \sum _n
t^n{\cal P}_n(\tilde \partial )\tau (T)
\eeq
where  ${\cal P}_n(\tilde \partial )$  are the Schur polynomials and symbol
$\tilde \partial $  represent the vector with components  $(-
N\partial /\partial T_1$, $- {1\over 2}N\partial /\partial T_2$ , ...).
Therefore from
eqs.(2.38), (2.39) and (3.24) one can deduce

$$
 - N{\partial \over \partial T_1}\tau (T) =
\partial _t\tau ^G_N[t|M]\left.\right|
_{t=0}
 = - N\sum _i \mu _i\cdot \tau (T) +
{1\over \Delta (\mu )}\lim _{t \to 0}
\partial _t{\rm det} \ \phi _i(\mu _j,t)\hbox{  ,}
$$
where  $\phi _i(\mu _j,t)$  are defined by eq.(2.39). From this expression it
follows that

\beq
{\partial \over \partial T_1}\tau (T) =
\sum \mu _i\cdot \tau (T) - \hat \tau (T)
\eeq
where  $\hat \tau (T)$  denotes some new $\tau $-function which is obtained
from the given  $\tau (T)$  by the shift of the last row in the determinant:
$\phi _N(\mu _j) \rightarrow  \phi _{N+1}(\mu _j)$. Eq.(3.25) is nothing but
eq.(3.13)\footnote{Let us emphasize that explicit form of {\it r.h.s.} of
(3.25) does not
depend on the choice of basis. Unlike this, the manifest form of
derivatives over $T_m$, with $m>1$, should deal with concrete choice of basis.
It is noteworthy to
remark that, for $m \leq K$, this manifest form coincides for
the canonical basis and our basis (2.18).}.

\subsection{Universal string equation}

The string equation, as implied by (3.23), is by definition:

\beq
{\partial \over \partial T_1}
{{\cal L}^{\{{\cal V}\}}_{-1}\tau ^{\{{\cal V}\}}\over \tau ^{\{{\cal V}\}}} =
0\ .
\eeq
If we substitute explicit expression (3.12) of the  ${\cal L}_{-1}$-operator
the string equation acquires the form:

\beq
\sum _{n\geq -1} Tr [{1\over {\cal V}''(M)M^{n+1}}]
{\partial ^2log\tau \over \partial T_1\partial T_n} = u\hbox{  .}
\eeq
Here we introduced a conventional notation  $u \equiv
\partial ^2log\tau /\partial T^2_1$ and also defined new  $T_0$- and
$T_{-1}$-derivatives, so that  $\partial log\tau /\partial T_0 \equiv  0$ ,
$\partial log\tau /\partial T_{-1} \equiv  T_1$. The derivation of (3.27)
involves taking $T_1$-derivatives of the objects

\beq
{\cal T} ^{\{{\cal V}\}}_n \equiv  Tr [{1\over {\cal V}''(M)M^{n+1}}].
\eeq
In order to get some impression about  $\partial {\cal T}_n/\partial T_1$, let
us imagine, that  ${\cal V}(M)$  is some polynomial of degree  $Q + 1$ , so
that  $[{\cal V}''(M)]^{-1} \sim  M^{1-Q}(1+v_1Q^{-1}+v_2Q^{-2}+...)$. Then
${\cal T}_n \sim  (Q+n)T_{Q+n} + v_1(Q+n+1)T_{Q+n+1} + v_2(Q+n+2)T_{Q+n+2} +
..$. . Therefore whenever  $Q+n > 1$ ($i.e$. $Q>0$, as it is already necessary
for the derivation of (3.23))

\beq
\partial {\cal T}_n/\partial T_1 = 0\hbox{,   for }\ n \geq  1.
\eeq
In order to derive (3.27) one also needs  $T_1$- derivative of the second item
in (3.12),

\beq
{1\over 2}\sum _{i,j} {1\over {\cal V}''(\mu _i){\cal V}''(\mu _j)}
{{\cal V}''(\mu _i)-{\cal V}''(\mu _j)\over \mu _i - \mu _j}\hbox{ .}
\eeq
Under the same assumptions about  ${\cal V}(\mu )$  the second ratio in this
sum is a polynomial in  $\mu$'s  of degree  $Q - 2$ . The only contribution to
(3.30), which contains pure factors  $\mu ^{-1}_i$ or  $\mu ^{-1}_j$ , is:
$\displaystyle {{1\over 2}\sum _{i,j}
{1\over {\cal V}''(\mu _i){\cal V}''(\mu _j)}\left\lbrace {{\cal V}''(\mu _i)
\over \mu _i} + {{\cal V}''(\mu _j)\over \mu  _j}\right\rbrace}  =
T_1{\cal T}_{-1}$,  others are expanded as bilinear series in  $T_mT_n$ with
$m$, $n > 1$. Therefore, $T_1$-derivative of (3.30) (if $Q > 2)$ is just
${\cal T}_1$ and is described by the  $n = 0$  term in (3.27).

So, we derived the universal string equation (3.27) in the form:

\beq
\sum _{n\geq -1} {\cal T}^{\{{\cal V}\}}_n
{\partial ^2log\tau ^{\{{\cal V}\}}\over \partial T_1\partial T_n} = u
\hbox{ . }
\eeq
If potential  ${\cal V}(X) = const\cdot X^{K+1}$ , the variables  ${\cal T}_n =
{n+K\over K} T_{n+K}$ , and we return to familiar relation \cite{GM90b}:

\beq
{1\over K}
\sum _{{n\geq -1}\atop {n \ne 0\ mod\ K}}
(n+K)T_{n+K} {\partial ^2log\tau ^{\{K\}}\over \partial T_1\partial T_n} =
u\hbox{  .}
\eeq
Note that even if we restore the factor (1.22), describing the maximal
possible  $T_{nK}$-dependence of  $\tau ^{\{K\}}$ it drops out of the string
equation (3.32) after taking the $\partial /\partial T_1$-derivative.

Our next task is reformulation of (3.31) in terms of pseudo-differential
operators and in the form of bilinear relation. If

\beq
L = \partial  + \sum ^\infty _i u_{i+1}\partial ^{-i}\hbox{ }
\ \ \ (\ \partial \equiv {\partial \over \partial T_1} \equiv {\partial \over
\partial x}\ )\ ,
\eeq
then \cite{DJKM83}

\beq
{\partial ^2log\tau \over \partial T_1\partial T_n} = (L^n)_{-1}\hbox{ .}
\eeq
For  $n = - 1$, 0, 1  we have  $(L^{-1})_{-1} = 1$ , $(L^0)_{-1} = 0$,
$(L)_{-1}
= u_2 \equiv u$  in accordance with (3.27). $K$-reduction is associated with
the
condition

\beq
(L^K)_- = 0
\eeq
(this guarantees, that items with  $n = 0$ $mod$ $K$ do not appear in (3.32)).
Rewritten in these terms, eq.(3.31) turns into:

\beq
Tr {1\over {\cal V}''(M)} \sum _{n\geq -1} M^{-n-1}(L^n)_{-1} =
u\hbox{  .}
\eeq
Recall now, that in the framework of the dressing formalism \cite{DJKM83} $L =
K\partial K^{-1}$,  $L^n = K\partial ^nK^{-1}$. The eigenfunctions of $L$ can
be defined as  $(Ke^{zx})$, where  brackets denote
that operator  $K$  acts on  $e^{zx}$ ($i.e$.
$(Ke^{zx})$  is a function of  $z$
, not an operator). Baker-Akhiezer function in these terms is given by
\cite{DJKM83}
$\Psi (z|T_n) = e^{\sum T_nz^n}(Ke^{zx})$, while its conjugate
$\Psi ^\ast (z|T_n) = e^{-\sum T_nz^n}[(K^{-1})^\ast e^{-zx}]$ . These
definitions are useful for us, since \cite{DJKM83}

\beq
(L^n)_{-1} = res_z z^n\Psi ^\ast (z)\Psi (z).
\eeq
We assume that contour integral over  $z$ , implicit in the definition of
$res_z$, is around zero. If (3.37) is substituted into (3.36), we obtain:

\beq
u= - res_zTr {M\over {\cal V}''(M)}{\Psi ^\ast (z)\Psi (z)\over z(zI - M)}\ .
\eeq
Since  $\Psi (z|T_n) = e^{\sum T_nz^n}\displaystyle {
{\tau (T_n-z^{-n}/n)\over \tau (T_n)}}$
and  $\Psi ^\ast (z|T_n) =
e^{-\sum T_nz^n}\displaystyle {
{\tau (T_n+z^{-n}/n)\over \tau (T_n)}}$,  the product
$\ \Psi ^\ast (z)\Psi (z) \sim  1 + {\cal O}(1/z)$ (as $z \to \infty$)
and the contour integral over
$z$ , when pulled to infinity, picks up the contributions from eigenvalues of
$M$ .
Finally we obtain our universal
string equation in the form of a bilinear relation for Baker-Akhiezer
functions:

\beq
\sum _i {\Psi ^\ast (\mu _i)\Psi (\mu _i)\over {\cal V}\ ''(\mu _i)} + u=0
\hbox{
.}
\eeq

Comparison of eqs.(2.46) and (3.39) enables us to derive a useful
representation for $u$-function:

\beq
u = -\sum _{a,b}{\partial ^2\over \partial \lambda _a\partial \lambda _b} \log
{\tau \over {\cal N}}\hbox{  .}
\eeq
Note that eqs.(2.45) and (2.46) which follow from the general
formula (2.44), provide the connection between residues of the Baker-Akhiezer
function and time derivatives of the $\tau $-function without reference to the
pseudo-differential operator
language. Indeed, it can be easily understood if one
substitutes eq.(3.25) into eq.(2.45).

\subsection{Discussion}

In this section 3 we derived the universal
${\cal L}^{\{\cal V\}}_{-1}$-constraint (1.25)
on partition function $Z^{\{{\cal V}\}}$.
Its simplest implication, the universal
string equation, was considered in sect.3.3. String equation is nothing but
$T_1$-derivative of this ${\cal L}_{-1}$-constraint.
Remarkably enough, it appears to be
much simpler than the constraint itself (compare (3.12) and (3.40)).

However, string equation is not the only implication of (1.23). Another one
should be the entire tower (1.29) of Virasoro and ${\cal W}$-constraints
imposed on
$Z^{\{{\cal V}\}}$.
In this subsection we discuss three different ways to derive these
remaining relations for any ${\cal V}(X)$
in the GKM. Being technically different,
they emphasize different (though related) properties of partition function, and
therefore each of the three deserves careful investigation, which is, however,
beyond the scope of this paper.

\bigskip
{\it KP-hierarchy approach.} This one is the most popular in the literature
(see
\cite{FKN91,DVV91a}). The idea is that ${\cal L}_{-1}$-constraint, when
imposed on KP
$\tau $-function, automatically implies all other constraints. In other words,
${\cal L}_{-1}$-constraint
is enough to fix the point in Grassmannian with which the
KP $\tau $-function is associated and, thus, it fixes this $\tau $-function
completely. It would be interesting to find out explicit dependence of
the point in Grassmannian on the potential ${\cal V}(X)$ in
GKM. Even more interesting
would be any alternative description of the entire subset $U_\infty $ in
Grassmannian, associated with GKM with arbitrary ${\cal V}(X)$,
and any reasonable
parametrization of $U_\infty $, which would be an alternative parametrization
of the space of potentials. Since ${\cal V}(X)$
can be changed smoothly, $U_\infty $
should be a kind of a manifold, lying at infinity of the universal module space
(if the latter is embedded into Sato's Grassmannian), and surely possess some
amusing properties. Unfortunately, the already existing papers on the subject
(of course, they concern conventional (multi)matrix models rather than GKM)
either emphasize implications of $K$-reduction, like \cite{KS91},
or rely upon the
formalism of pseudo-differential operators, like
\cite{FKN91,DVV91a}, and thus need be
translated into Grassmannian language.

A separate question in the framework of this approach is why should the
constraints be always associated with ${\cal W}_\infty $-algebra
(as they actually
are). Technically this is more or less obvious whenever pseudo-differential
operators are used. In KP-Grassmannian language, the crucial ingredient should
be the ${\cal W}_\infty $-covariance of
KP-hierarchy, discovered in \cite{OS86,GO89,Sem89}. The
constraints of interest involve operators from some Borel subalgebra
${\cal W}^+_\infty $
of this ${\cal W}_\infty $, and the choice of subalgebra depends on the
choice of potential.

Reductions of GKM to (double-scaling limit of) multimatrix models, $i.e$. the
what happens if ${\cal V}(X) \sim  X^{K+1}$, in Grassmannian language should be
attributed to
intersections of $U_\infty $ and submanifolds $Gr^{\{K\}}$ in Sato's
Grassmannian, associated with $K$-reductions. Somehow at these points a closed
Zamolodchikov's ${\cal W}_K$-subalgebra emerges from entire ${\cal W}_\infty $.
It is
certainly interesting to see, how this happens and how the Lie algebra
structure is broken. It is also unclear, whether such phenomenon takes place
only at $U_\infty \cap Gr^{\{K\}}$ or everywhere in $Gr^{\{K\}}.$

One more important question is what is the relation between the just discussed
${\cal W}_\infty $-algebra of \cite{OS86,GO89,Sem89} and another
${\cal W}_\infty $, which is
presumably
relevant in $c=1$ models, and is naively generated by operators, which involve
finite-differences instead of derivatives in the free-field representation. We
use this chance to note, that the problem of how $c=1$ models are included into
framework of GKM is still open, and the invariant description of the
$U_\infty $ subset in Grassmannian would be also helpful for its resolution.

\bigskip
{\it Straightforward derivation of ${\cal L}$- and ${\cal W}$-constraints.}
This approach is
just the straightforward generalization of the what was done in sect.3.2 in the
derivation of ${\cal L}^{\{{\cal V}\}}_{-1}$-constraint
from the knowledge of explicit
expression (2.2) for $Z^{\{{\cal V}\}}$. We sketch here several steps of the
derivation for the Virasoro case. The idea is to apply the operator
$Tr\Lambda ^{m+1}{\partial \over \partial \Lambda _{tr}}$ to (2.2).
On one hand this is equal to

\beq
\sum _{n\geq 1} Tr\left( {({\cal V}'(M))^{m+1}\over {\cal
V}''(M)M^{n+1}}\right)
{\partial \over \partial T_n} Z^{\{{\cal V}\}},
\eeq
on the other hand, it can be explicitly evaluated. Expression (3.41) is very
close to ${\cal L}^{\{{\cal V}\}}_mZ^{\{{\cal V}\}}$. The main difference is
that
the sum in (3.41)
goes from $n=1$ and for small enough
$n$ $Tr\left(\displaystyle {
{({\cal V}'(M))^{m+1}\over {\cal V}''(M)M^{n+1}}}\right) $
contains powers of
{\it positive} powers of $M$, which can not be expressed through $T$-variables.
For ${\cal V} \sim  X^{K+1}$ this happens for $n\leq Km$. In order to get
rid of these
positive powers of $M$ one should use the analogues of eq. (3.14), saying that

\beq
{\partial\over \partial T_k} \log \ Z^{\{{\cal V}\}} \sim  Tr\ M^k
+ ..\hbox{. .}
\eeq
This is the origin of conventional $\partial ^2/dT_adT_b$-terms in operators
${\cal L}^{\{K\}}$. As to generalization of the shift (1.17), it results in
additional contribution of the form

$$
- \sum _{n\geq 1} \oint{[{\cal V}'(\mu )]^{m+1}\over \mu ^n}{\partial \over
\partial T_n}
$$
in the definition of ${\cal L}^{\{V\}}_m$ - operators.
All extra terms cancel, giving rise to the proper universal
constraints of the form ${\cal L}^{\{{\cal V}\}}_mZ^{\{{\cal V}\}} =
0$, $m\geq -1.$

${\cal W}$-constraints can be (at least in principle) derived in the same
manner,
though actual calculations are increasingly sophisticated.

\bigskip
{\it Implications of Ward identities.} This third approach exploits the
fundamental Ward-identity (1.23). Technically its main difference from the
previous approach is that the operators $Tr{\cal V}'({\partial \over \partial
\Lambda _{tr}})$,
non-linear in $\displaystyle {\partial /\partial \Lambda _{tr}}$,
are involved. Conceptually this approach is
different, since it does not exploit explicitly any integrable features of
partition function. We shall discuss this method in more details in the next
section 4.

\section{From Ward identities to  $W$-constraints}
\setcounter{equation}{0}
\subsection{General discussion}

This section is devoted to the derivation (unfortunately, incomplete) of the
entire set of Virasoro and  ${\cal W}$- constraints in GKM. The role of such
constraints in the study of any matrix model is two-fold. First of all, they
can be considered as complete set of differential equations, which specify
partition
function of the model as a function of time-variables. Then
this set of equations implies, among other things,
that the solution is KP $\tau $-function
(and sometimes a reduced $\tau ^{\{K\}}$-function). The first application,
discussed
 in details in sect.3 above, is
reasonable if one already knows about the integrable structure. Then partition
function is identified with {\it some} (reduced) KP $\tau $-function $(i.e$.
evaluated at {\it some} point of Grassmannian), and the role  of Virasoro
constraint is to specify this $\tau $-function $(i.e$. fix the point in
Grassmannian). In this second type of circumstances it is enough to have a
single constraint, namely  ${\cal L} _{-1}Z = 0$ , --- all other constraints,
if imposed on KP $\tau $-function or  $\tau ^{\{K\}}$ , are deducible
corollaries of this one \cite{FKN91,DVV91a}. With identification of
$Z^{\{{{\cal V}} \}}$
with a $\tau $-function in sect.2 and the proof of universal  ${\cal L}
^{\{{\cal V} \}}_{-1}$- constraint in sect.3 we exhausted this line of
reasoning. The subject of this section is to concentrate instead on another
approach and ignore almost all what we already studied about
integrable structure of GKM. Then the constraints can be considered as a
{\it complete} set of differential equations, which specify the partition
function of the model as a function of time-variables. Among other things,
the set of equations implies that the solution is KP $\tau$-function (and
sometimes a reduced $\tau ^{\{K\}}$-function). To be more concrete,
we shall investigate direct
corollaries of Ward identity (1.23),

\beq
\{Tr\ \epsilon (\Lambda )[{\cal V}\ '(\partial /\partial \Lambda _{tr}) -
\Lambda ]\} {\cal F}_N[\Lambda ] = 0\hbox{.}
\eeq
We shall, however, discuss only specific potentials,  ${\cal V} (X) =
const\cdot X^{K+1}$ , and use explicitly the fact that partition function is
independent of all  $T_{nK}.$

The main problem with the implications of Ward identity (4.1) is that they
acquire the form of conventional Virasoro or  ${\cal W}$-constraints
only in the limit of  $N =
\infty $. The reliable results from our point of view, should, however, be
$N$-independent. But as soon as  $N$  is finite, the complete set of
{\it independent} Ward identities (4.1) is also finite. Remarkably enough they
can still be expressed through generators of ${\cal W}$-algebras,
but the constraints
arise as a {\it finite}
number of vanishing conditions for infinite linear
combinations of  ${{\cal W}}$-operators, acting on partition function. As  $N
\rightarrow  \infty $ , the number of such conditions tends to infinity,
implying
that {\it every} item in linear combinations vanishes by itself. These
items have exactly the form of conventional Virasoro constraints in the
particular case of  $K =2$, while for  $K = 3$  they rather look like
eqs.(1.26), and generalization of (1.26) for  $K > 3$  is more or less obvious.
In any case the honest statement is that any solution to the  ${\cal W}$
-constraints in the conventional form (1.29) does satisfy the equations, which
follow from (4.1), for any  $N$  (to make  $N$  finite one should take all but
the first  $N$  eigenvalues of matrix  $M$  to infinity, see sect.2.2). In this
sense transition from finite to infinite  $N$  is smooth. However, inverse can
be true, $i.e$. (4.1) can imply (1.29) only as  $N = \infty $. Moreover, since
(1.26), which is actually implied by (4.1) as $N = \infty $ , is not quite
identical to (1.29), one needs also to
rely upon the (very plausible) assumption
that solutions to (1.26) and (1.29) are unique (up to inessential constant
factors) and thus coincide.

Such approach has been already applied in \cite{MMM91b}
to the study of Kontsevich
model itself, $i.e$. for the case of  $K =2:  {\cal V} _2(X) = X^3/3$. Our
purpose now is to extend this consideration to other potentials  ${\cal V}
_K(X) = X^{K+1}/(K+1)$. However, while the structure of the answers is very
clear (this is obvious from (1.26)),
actual calculations are very tedious. Therefore
we restrict our presentation below only to the first non-trivial case of $K =
3$ . It is considered in sect.4.3. Before, in sect.4.2 we reproduce from
\cite{MMM91b} the derivation of Virasoro constraints in the case of  $K = 2.$

\subsection{Virasoro constraints in Kontsevich model $(K = 2)$}

{\it The problem.}
Our purpose in this section is to prove the identity

\beq
{1\over {\cal F}} tr(\epsilon _p {\partial ^2\over \partial \Lambda ^2_{tr}} -
\epsilon _p\Lambda ){\cal F} = {1\over Z}
\sum _{n\geq -1} {\cal L}_nZ \hbox{   } tr(\epsilon _p\Lambda ^{-n-2})
\eeq
for

\beq
{\cal F}^{\{2\}}\{\Lambda \} \equiv  \int   DX\ \exp (- trX^3/3 + tr\Lambda X)
= C[\sqrt{\Lambda }] \exp ({2\over 3}tr\Lambda ^{3/2})
Z^{\{2\}}_{tg}(T_m)\hbox{ , }\footnote{To avoid any comparison note
that in original version of \cite{MMM91b}
there were wrong signs
in the exponential in (4.3) and in the
shift of times (1.17) (normalization of times are also different
).}\ \ \ m\ \ -\ \ odd
\eeq
with

\beq
C[\sqrt{\Lambda }] = {\rm det} (\sqrt{\Lambda }\otimes I +
I\otimes \sqrt{\Lambda })^{-{1\over2}}
\eeq
and

\bea
{\cal L}^{\{2\}} _n = {1\over 2}\sum _{{k\geq \delta_{n+1,0} }\atop {k\ odd}}
kT_k {\partial \over \partial T_{k+2n}} + {1\over 4}
\sum _{^{a+b=2n}_{a,b\geq 0\ ;\ a,b\ odd}}
{\partial ^2\over \partial T_a\partial T_b} + \nn \\
+ \delta _{n+1,0}\cdot {T^2_1\over 4} + \delta _{n,0}\cdot {1\over 16}
- {\partial \over \partial T_{2n+3}}\hbox{ . }
\eea
Below in this section we omit label $\{2\}$ and do not indicate explicitly
the fact that all sums run over only odd times.

  While (4.2)
is valid {\it for any size of the matrix $\Lambda $}, in the limit of
infinitely large $\Lambda $ ($N \to \infty$) we can insist that
all the quantities

\beq
tr(\epsilon _p\Lambda ^{-n-2})
\eeq
$(e.g$. $tr\Lambda ^{p-n-2}$ for $\epsilon _p = \Lambda ^p)$
become algebraically independent, so that eq.
(4.2) and (4.1) imply that

\beq
{\cal L} _nZ\{T\} = 0\hbox{, }    n \geq  -1 \ .
\eeq

\bigskip
{\it Method.}
Note that ${\cal F} \{\Lambda \}$ in (4.3), which
we have to differentiate in order to prove (4.2),
depends only upon eigenvalues
$\{\lambda _k\}$ of the matrix $\Lambda $. Therefore, it is natural to consider
eq.(4.2) at the diagonal point $\Lambda _{ij}=0$, $i\neq j$. The only
``non-diagonal" piece of (4.2) which survives at this point is proportional to

\beq
\left.{\partial ^2\lambda _k\over \partial \Lambda _{ij}\partial \Lambda _{ji}}
\right| _{\Lambda _{mn}=0\hbox{, } m\neq n} =
{\delta _{ki}-\delta _{kj}\over \lambda _i-\lambda _j}\hbox{
for }\ i\neq j.
\eeq
Eq.(4.8) is nothing but a familiar formula for the second order correction to
 Hamiltonian
eigenvalues in ordinary quantum-mechanical perturbation theory. It can be
easily derived from the variation of determinant formula:

\begin{eqnarray}
\delta log({\rm det} \ \Lambda ) = tr {1\over \Lambda } \delta \Lambda  -
{1\over 2}
tr( {1\over \Lambda } \delta \Lambda  {1\over \Lambda } \delta \Lambda )
 + \ldots \hbox{ .}
\end{eqnarray}
For diagonal $\Lambda _{ij}= \lambda _{i}\delta _{ij}$, but, generically,
non-diagonal $\delta \Lambda _{ij}$, this equation gives

\begin{eqnarray*}
\sum  _k {\delta \lambda _k\over \lambda _k} = - {1\over 2} \sum _{i\neq j}
{\delta \Lambda _{ij}\delta \Lambda _{ji}\over \lambda _i\lambda _j} =
{1\over 2} \sum _{i\neq j} \left( {1\over \lambda _i} -
{1\over \lambda _j}\right)
{\delta \Lambda _{ij}\delta \Lambda _{ji}\over \lambda _i-\lambda _j} +
\ldots \hbox{ ,}
\end{eqnarray*}
which proves (4.8).

\bigskip
{\it Proof.}
Now we shall turn directly to the proof of (4.2).
Since $\epsilon _p$ is assumed
to be a function of $\Lambda $, it can be, in fact, treated as a function of
eigenvalues $\lambda _i$. After that, (4.2) can be rewritten in the following
way:

\beq
{e^{-{2\over 3}tr\Lambda ^{3/2}}\over C(\sqrt \Lambda )Z\{T\}}\left[ tr\
\epsilon _p\{{\partial ^2\over \partial \Lambda ^2} -
\Lambda \}\right]  C(\sqrt \Lambda ) e^{
{2\over 3}tr\Lambda ^{3/2}}Z\{T\} =
\eeq

\beq
= {1\over Z}\sum _{a,b\geq 0}
{\partial ^2Z\over \partial T_a\partial T_b}
\sum  _i \epsilon _p(\lambda _i)
{\partial T_a\over \partial \lambda _i}\cdot {\partial T_b\over
\partial \lambda _i} +
\eeq

\bea
+ {1\over Z} \sum _{n\geq 0}
{\partial Z\over \partial T_n} \left[ \sum _{i,j}
\epsilon _p(\lambda _i)
{\partial ^2T_n\over \partial \Lambda _{ij}\partial \Lambda _{ji}} + 2 \sum  _i
\epsilon _p(\lambda _i) {\partial T_n\over \partial \lambda _i}
{\partial logC\over \partial \lambda _i} +\right. \nn \\
+ \left. 2 \sum  _i \epsilon _p(\lambda _i) {\partial T_n\over \partial
\lambda _i}
\left( {2\over 3}\right)  {\partial \over \partial \lambda _i}
tr\Lambda ^{3/2} \right]  +
\eea

\beq
+ \left[
 \sum  _i
\epsilon _p(\lambda _i)
\left( {\partial \over \partial \lambda _i}\left(
{2\over 3}\right)  tr\Lambda ^{3/2}\right) ^2 \right.
-  \sum  _i \lambda _i \epsilon _p(\lambda _i) +
\eeq

\beq
+ \sum _{i,j}\epsilon _p(\lambda _i)
\left( {\partial ^2\over \partial \Lambda _{ij}\partial \Lambda _{ji}}
\left( {2\over 3}\right)  tr\Lambda ^{3/2}\right)  +
\eeq

\beq
+ 2 \sum  _i \epsilon _p(\lambda _i) \left(
{2\over 3}\right)
{\partial tr\Lambda ^{3/2}\over \partial \lambda _i}
{\partial logC\over \partial \lambda _i} +
\eeq

\beq
+ \left. {1\over C} \sum _{i,j}\epsilon _p(\lambda _i)
{\partial ^2C\over \partial \Lambda _{ij}\partial \Lambda _{ji}} \right]
\eeq
with  $tr\Lambda ^{3/2} = \sum  _k \lambda ^{3/2}_k$ and

\beq
C = \prod _{i,j}(\sqrt{\lambda }_i + \sqrt{\lambda }_j)^{-1/2}.
\eeq

First, since

\beq
{\partial T_n\over \partial \lambda _i} = - \lambda ^{-n-{3\over 2}}_i\ ,
\eeq
it is easy to notice that the term with the second derivatives (4.11) can be
immediately written in the desired form:

\bea
(4.11) ={1\over 4}\sum _{n\geq -1}
 \sum  _i \epsilon _p(\lambda _i) \lambda ^{-n-2}_{i}
\sum _{a+b=2n}
{\partial ^2Z\over \partial T_a\partial T_b}= \nn \\
={1\over 4}\sum _{n\geq -1
}tr\{\epsilon _p(\lambda _i)\Lambda ^{-n-2}\}\sum _{a+b=2n}{\partial ^2Z
\over \partial T_a\partial T_b}\hbox{.}
\eea
For the first-derivative terms with the help of (4.17),

\beq
{\partial logC\over \partial \lambda _i} = - {1\over 2\sqrt \lambda _i} \sum
_j {1\over \sqrt \lambda _i+\sqrt \lambda _j}
\eeq

\beq
{\partial \over \partial \lambda _i} tr\Lambda ^{3/2} = {3\over 2}
\sqrt{\lambda }_i
\eeq
and

\beq
{\partial ^2T_n\over \partial \Lambda _{ij}\partial \Lambda _{ji}} = \sum  _k
{\partial ^2\lambda _k\over \partial \Lambda _{ij}\partial \Lambda _{ji}}
{\partial T_n\over \partial \lambda _k} +
{\partial ^2T_n\over \partial \lambda^2_i}
 \delta _{ij}\hbox{ .}
\eeq
Then for (4.12) we have

\bea
{1\over 2Z} \sum _{n\geq 0} {\partial Z\over \partial T_n}
\left[ \sum _{i,j} \epsilon _p(\lambda _i) \left\{{\lambda_ i^{n-3/2}
\lambda_j^{n-3/2}\over \sqrt \lambda _i+\sqrt \lambda_j}
+\sum _{a+b=2(n+1)}(\sqrt{\lambda }_i)^a(\sqrt{\lambda }_j)^b
+\right. \right. \nn \\
+ \left. \left. \lambda ^{-n-2}_i {1\over \sqrt \lambda _i+\sqrt \lambda _j}
 \right\rbrace  - 2 \sum  _i \epsilon _p(\lambda _i)
\lambda ^{-n-1}_i  \right]  = \nn \\
= {1\over 2Z} \sum _{n\geq 0}{\partial Z\over \partial T_n} \left[ \sum _{i,j}
\sum ^{n-1}_{a=0}\epsilon _p(\lambda _i) \lambda ^{-n-2+a}_i
\lambda ^{-a-1/2}_j - 2\sum  _i
\epsilon _p(\lambda _i)\lambda _i^{-n-1}\right]  = \nn \\
= \left.{1\over 2Z} \sum _{n\geq -1} \sum  _i \epsilon _p(\lambda _i)
\lambda ^{-n-2}_{i} \sum _{k\geq \delta _{n+1,0}}\left( \sum  _j
\lambda ^{-k-1/2}_j -  2\delta _{k,3}\right)
{\partial Z\over \partial T_{2n+k}} \right]  =  \nn \\
= {1\over Z} \sum _{n\geq -1}tr(\epsilon _p\Lambda ^{-n-2})\left\{
{1\over 2}\sum _{k\geq
 \delta _{n+1,0}}
k T_k {\partial Z\over \partial T_{2n+k}} - {\partial Z \over
\partial T_{2n+3}}\right\}.
\eea
The remaining part contains terms proportional to $Z$ itself which need a bit
more care. First, it is easy to notice that two items in (4.13)
just cancel each
other, so the (4.13) gives no contribution to the final result. For (4.14)
we have

\beq
  \sum  _i \epsilon _p(\lambda _i)
{\partial ^2tr\Lambda ^{3/2}\over \partial \lambda^2_i
} + \sum _{i,j}\epsilon _p(\lambda _i)
{\partial tr\Lambda ^{3/2}\over \partial \lambda _k}
{\partial ^2\lambda _k\over \partial \Lambda _{ij}\partial \Lambda _{ji}}
\hbox{ . }
\eeq
Using (4.21) and (4.8) it can be transformed into

\bea
 {1\over 2} \sum  _i \epsilon _p(\lambda _i)
\lambda ^{-1/2}_i + \sum _{i\neq j}\epsilon _p(\lambda _i)
{\sqrt{\lambda }_i-\sqrt{\lambda }_j\over \lambda _i-\lambda _j}
  = \nn \\
=  \sum _{i,j}\epsilon _p(\lambda _i)
{1\over \sqrt \lambda _i+\sqrt \lambda _j}
\eea
and this cancels (4.15) (transformed with the help of (4.20), (4.21)).
Thus, the only contribution is (4.16), which gives

\beq
\new
\begin{array}{l}
{1\over C} {\partial ^2C\over \partial \Lambda _{ij}\partial \Lambda _{ji}}
= \sum  _k
{\partial ^2\lambda _k\over \partial \Lambda _{ij}\partial \Lambda _{ji}}
{\partial logC\over \partial \lambda _k} + {1\over C}
{\partial ^2C\over \partial \lambda^2_i
} \delta _{ij} = \nn \\
= \sum  _k {1\over 2(\lambda _i-\lambda _j)}
\left[ {1\over \sqrt \lambda _j(\sqrt \lambda _j+\sqrt \lambda _k)} -
{1\over \sqrt \lambda _i(\sqrt \lambda _i+\sqrt \lambda _k)}\right]  (1 -
\delta _{ij}) + \nn \\
+ \delta _{ij}
\left[ {\partial ^2\hbox{logC}\over \partial \lambda^2_i
} + \left( {\partial logC\over \partial \lambda _i}\right) ^2\right] .
\end{array}
\eeq

Now, using (4.8) and (4.20) we obtain for (4.16):

\begin{eqnarray}
 & & \left\lbrace {1\over 16} \sum  _i \epsilon _p(\lambda _i)
\lambda ^{-2}_i +\right. \nn\\
& + &  {1\over 4} \sum _{i,j}\epsilon _p(\lambda _i) \lambda ^{-1}_i
\left( {1\over \sqrt \lambda _i(\sqrt \lambda _i+\sqrt \lambda _j)} +
{1\over (\sqrt \lambda _i+\sqrt \lambda _j)^2}\right)  + \nn \\
& + & {1\over 4} \sum _{i,j,k}\epsilon _p(\lambda _i) \lambda ^{-1}_i
{1\ \ \ \ \ \ \ \ 1\over (\sqrt \lambda _i+\sqrt \lambda _j)(\sqrt \lambda _i+
\sqrt \lambda _k)} + \nn \\
& + & {1\over 2}\sum _{i\neq j,k}{\epsilon _p(\lambda _i)\over
 \lambda _i-\lambda _j
} \left[ {1\over \sqrt \lambda _j(\sqrt \lambda _j+\sqrt \lambda _k)} -
{1\over \sqrt \lambda _i(\sqrt \lambda _i+\sqrt \lambda _k)}\right] \left.
\right\rbrace  =  \nn \\
& = & {5\over 16} \sum  _i \epsilon _p(\lambda _i) \lambda ^{-2}_i +
\sum _{i\neq j}\left( {1\over 4} \epsilon _p(\lambda _i)
\lambda ^{-1}_i\lambda ^{-1}_j + {1\over 2} \epsilon _p(\lambda _i)
\lambda ^{-3/2}_i\lambda ^{-3/2}_j\right)  + \nn \\
& + & \sum _{i\neq j\neq k}\epsilon _p(\lambda _i)\left( {1\over
 4\lambda _i(\sqrt
\lambda _i+\sqrt \lambda _j)(\sqrt \lambda _i+\sqrt \lambda _k)} +\right. \nn
\\
& + & {1\over 2(\lambda _i-\lambda _j)}\left[\left.{1\over \sqrt
 \lambda _j(\sqrt \lambda
_j+\sqrt \lambda _k)} -
{1\over
 \sqrt \lambda _j(\sqrt \lambda _i+\sqrt \lambda _k)}\right] \right)  = \nn \\
& = & {5\over 16} \sum  _i \epsilon _p(\lambda _i) \lambda ^{-2}_i +
\sum _{i\neq j}\left( {1\over 4} \epsilon _p(\lambda _i)
\lambda ^{-1}_i\lambda ^{-1}_j + {1\over 2} \epsilon _p(\lambda _i)
\lambda ^{-3/2}_i\lambda ^{-3/2}_j\right)  + \nn \\
& + & {1\over 4}\sum _{i\neq j\neq k}\epsilon (\lambda )
\lambda ^{-1}_i\lambda ^{-1/2}_j\lambda ^{-1/2}_k.
\eea
In the last transformation we used the fact that

\beq
\new
\begin{array}{l}
\sum _{j\neq k}
{1\over (\lambda _i-\lambda _j)}\left[ {1\over \sqrt \lambda _j(\sqrt \lambda _
j+\sqrt \lambda _k)} -
{1\over \sqrt \lambda _j(\sqrt \lambda _i+\sqrt \lambda _k)}\right]  = \\
= {1\over 4} \sum _{j\neq k}
\left\lbrace {1\over (\lambda _i-\lambda _j)}\left[ {1\over \sqrt \lambda _j(
\sqrt \lambda _j+\sqrt \lambda _k)} -
{1\over \sqrt \lambda _j(\sqrt \lambda _i+\sqrt \lambda _k)}\right]
 + (j\leftarrow \rightarrow k)\right\rbrace \hbox{.}
\end{array}
\eeq

Finally, (4.27) can be rewritten as

\bea
{1\over 16} \sum  _i \epsilon _p(\lambda _i) \lambda ^{-2}_i + {1\over 4} \sum
_i \epsilon _p(\lambda _i) \lambda ^{-1}_i \sum _{j,k}
\lambda ^{-1/2}_j\lambda ^{-1/2}_k = \nn \\
= \sum  _n \left( \sum  _i \epsilon _p(\lambda _i)
\lambda ^{-n-2}_i\right) \left\lbrace {1\over 16} \delta _{n,-1}\left[ 2 \sum
_j \lambda ^{-1/2}_j\right] ^2 + {1\over 16} \delta _{n,0}\right\rbrace .
\eea

Now taking together (4.19), (4.23) and (4.29) we obtain our main result:

\bea
{e^{-{2\over 3}tr\Lambda ^{3/2}}\over C(\sqrt \Lambda )Z\{T\}}\left[ tr\
\epsilon _p\{{\partial ^2\over \partial \Lambda ^2} -
\Lambda \}\right]  C(\sqrt \Lambda ) e^{
{2\over 3}tr\Lambda ^{3/2}}Z\{T\} = \nn \\
= {1\over Z} \sum _{n\geq -1}tr(\epsilon _p\Lambda ^{-n-2})
\left\lbrace {1\over 2}\sum _{k\geq \delta _{n+1,0}}kT_k
{\partial \over \partial T_{2n+k}}
+{1\over 4}\sum _{^{a+b=2n}_{a\geq 0,b\geq 0}}{\partial ^2\over \partial T_a
\partial T_b} +\right.\nn \\
+ \left.
{1\over 16} \delta _{n,0} + {1\over 4} \delta _{n+1,0} T^2_1 - {\partial
\over \partial T_{2n+3}}
\right\rbrace  Z(T) = 0.
\eea
This completes the derivation of (4.2) and, thus, of Virasoro constraints for
the case $K=2$.

\subsection{Example of $K = 3$}

In the case of generic  $K$  the analogue of the derivation, described in the
previous section, should involve the following steps.

--- Represent  ${\cal F} [\Lambda ]$  as

\beq
{\cal F} ^{\{K\}}[\Lambda ] = g_K[\Lambda ]Z^{\{K\}}(T_n)\hbox{,}
\eeq
with

\beq
g_K[\Lambda ] = {\Delta (M)\over \Delta (\Lambda )}\prod _i
[{\cal V}\ ''(\mu _i)^{-1/2} e^{(\mu _i{\cal V}\ '(\mu _i)-{\cal V} (\mu _i))}]
= {\Delta (\Lambda ^{1/K})\over \Delta (\Lambda )}\prod _i
[\lambda ^{-{K-1\over 2K}}_ie^{\alpha {K\over K+1}\lambda ^{1+1/K}_i}]\hbox{.}
\eeq
Parameter  $\alpha $  is introduced here for the sake of convenience, in fact,
$\alpha  = 1.$

--- Substitute this  ${\cal F} ^{\{K\}}[\Lambda ]$  into (4.1), which in the
particular case of  $\displaystyle {{\cal V} _K(X) = {X^{K+1}\over (K+1)}}$,
looks like

\beq
\{Tr\ \epsilon (\Lambda )[({\partial \over \partial \Lambda _{tr}})^K -
\alpha ^K\Lambda ]\} g_K[\Lambda ]Z^{\{K\}}(T_n) = 0.
\eeq
When  $\partial /\partial \Lambda _{tr}$ acts on $Z(T_n)$, the following rule
is applied:

\beq
{\partial Z\over \partial \Lambda _{tr}} = -{1\over K} \sum _{n \geq 1}
(n+K)T_{n+K}{\partial Z\over \partial T_n},
\eeq
and the sum at the $r.h.s$. goes over all positive  $n \neq  0$ mod K. As to
higher-order derivatives,  $\displaystyle {{\partial ^iZ\over
\partial \Lambda ^i_{tr}}}$ ,
they are defined with the help of relations like (4.8).

--- When all the  $\Lambda $-derivatives in (4.1) act on exponent in
$g[\Lambda ] = g[{\cal V}\ '(M)]$ , we get the term, which is equal to
${{\cal V}}\ '\displaystyle {({\partial Tr(M{{\cal V}}\ '(M)-{{\cal V}}
(M))\over \partial \Lambda _{tr}}) = {{\cal V}}\ '(M)}
= \Lambda $ and cancels
$\Lambda $-term in (4.1). The next contribution, when all but one of the
$\Lambda $-derivatives act on the exponent, vanishes. Actually this reflects
the fact that there are no  ${\cal W} ^{(1)}$- generators among the final
${\cal W}$-
constraints.

--- Perform a shift of variables

\beq
T_n\to \hat T_n = T_n - \alpha
{K\over n}\delta _{n,K+1}
\eeq
(this procedure doesn't change $\partial /
\partial T_n \to \partial /\partial \hat T_n$).

--- After all these substitutions the $l.h.s$. of eq.(4.33) acquires the form
of
an infinite series where every item is a product of
$Tr[\tilde \epsilon (M)M^{-p}]$  and a linear combination of generators of
${\cal W}_K$-algebra,
acting on  $Z^{\{K\}}(T_n)$. In the case of  $K = 3$  this
equation looks like

\beq
\new
\begin{array}{c}
{1\over 27}
Tr\left[\tilde \epsilon (M)M^{-3}\left\lbrace \sum _{n\geq -2}M^{-3n}
{\cal W} ^{(3)}_{3n} +\right.\right. \\
+ 9\sum _{n\geq -2}M^{-3n-1/3} \left\{ \sum  (3k-2)\hat T_{3k-2}
{\cal W} ^{(2)}_{3n+3k} +
\sum _{{a+b=3n}\atop {a,b\geq 0;\ n\geq -3}} {\partial \over \partial T_{3a+1}}
{\cal W} ^{(2)}_{3b-3}\right\}+ \\
\left. \left. + 9\sum _{n\geq -2}M^{-3n-2/3}\left\{ \sum  (3k-2)\hat T_{3k-2}
{\cal W} ^{(2)}_{3n+3k} +
\sum _{{a+b=3n}\atop {a,b\geq 0;\ n\geq -3}} {\partial \over \partial T_{3a+1}}
{\cal W} ^{(2)}_{3b-3}\right\}  \right\rbrace \right] Z^{\{3\}} = 0,
\end{array}
\eeq
(see (1.27),(1.28) for explicit expressions for ${\cal W}$-operators).
This relation is valid for {\it any} value of  $N$ , and in this sense is an
example of identity, which behaves smoothly in the limit  $N \rightarrow
\infty .$

--- If  $N = \infty $  all the quantities  $Tr\tilde \epsilon (M)M^{-p}$ with
given  $p$  but varying  $\tilde \epsilon (M)$ become independent, and (4.36)
may be said to imply (1.26).

In what follows we restrict ourselves to an illustrative calculation: we prove
that (4.36) holds for  $N = 1$  ($i.e$. when there are numbers instead of
matrices). Even this calculation is long enough. Additional technical details,
necessary to deal with matrices rather than ordinary numbers, are exhaustively
discussed in the previous section 4.2.

\bigskip
{\it Example.}
When (4.32) with  $K = 3$,  $N = 1$  is substituted into (4.33) and the rule
(4.34) is used, we obtain $(Z \equiv  Z^{\{3\}}_1):$

\beq
\new
\begin{array}{l}
0 = \tilde \epsilon (\mu )\left\lbrace \sum _{n\geq 1}
{\alpha ^2\over \mu ^{n+1}} {\partial Z\over \partial T_n} -
{\alpha \over \mu ^5} \left[ {1\over 3} \sum _{m,n} {1\over \mu ^{m+n}}
{\partial ^2Z\over \partial T_m\partial T_n} + \sum _n{n+4\over 3}
{1\over \mu ^n} {\partial Z\over \partial T_n} + {7\over 9} Z\right]
+\right.\nn \\
+  {1\over \mu ^3} \left[ \sum _{l,m,n} {1\over 27\mu ^{l+m+n}}
{\partial ^3Z\over \partial T_l\partial T_m\partial T_n} +
\sum _{m,n}{n+m+8\over 18}
{1\over \mu ^{m+n}} {\partial ^2Z\over \partial T_m\partial T_n}
+\right.\nn \\
+ \left. \left.
\sum _n{n^2+12n+39\over 27} {1\over \mu ^n} {\partial Z\over \partial T_n}
+ {28\over 27} Z \right] \right\rbrace
\end{array}
\eeq
Our purpose now is to compare this expression with the $l.h.s$. of (4.36). In
order to understand the structure of (4.36) let us note, that the first item
(with  ${\cal W} ^{(3)})$ in (4.36) contains the contribution
$\displaystyle {{1\over \mu ^{l+m+n}}
{\partial ^3Z\over \partial T_l\partial T_m\partial T_n}}$ ,
just the same as in
(4.37), but only under restriction  $l + m + n = 0$  $mod$ $3$. However, no
such
restriction is imposed in (4.37), and in order to restore the equivalence
between (4.36) and (4.37), one needs to add the terms
with  $\partial ({{\cal W}^{(2)}}\ Z)/\partial T$ to the $l.h.s.$ of (4.36),
which add the missing contributions
$\displaystyle {{1\over \mu ^{l+m+n}}
{\partial ^3Z\over \partial T_l\partial T_m\partial T_n}}$  with  $l + m + n =
1$, $2$ $mod$ $3$.

If we analyze the terms without  $\alpha $  in (4.37), it is important that
$\alpha $  appears also in the shift (4.35):

\beq
n\hat T^{\{3\}}_n = nT_n - 3\alpha \delta _{n,4},
\eeq
so that at the same time we substitute all  $\hat T$'s  by  $T$'s  in (4.36),
(1.26)--(1.28).
Then the term  $\displaystyle {{(K-1)(3K-1)(5K-1)...((2K-1)K - 1)\over
(2K)^K} Z
= {28\over 27}} Z$  in (4.37)
should be compared to the contributions without
$Z$-derivatives to (4.36). These come from the negative harmonics of
${\cal W} ^{(3)}$ and  ${\cal W} ^{(2)}$ operators. Namely,  so modified
${\cal W}
^{(3)}_{-6}$ contains  $(2T_2)^3 + 3(T_1)^2(4T_4) = \displaystyle
{{4\over \mu ^6}}$,
${\cal W} ^{(3)}_{-3}$ contains  $(T_1)^3 = \displaystyle {
{1\over \mu ^3}}$ , and there is no
constant term in ${\cal W} ^{(3)}_0$. As to Virasoro operators, there is
$\displaystyle {{1\over 6}}\cdot 2(T_1)(2T_2) = \displaystyle {
{1\over 3\mu ^3}}$ in  ${\cal W} ^{(2)}_{-3}$
and  $\displaystyle {{(K-1)(K+1)\over 24K} = {1\over 9}}$
in  ${\cal W} ^{(2)}_0$ . Also the
contribution of interest to   $\partial {\cal W} ^{(2)}_{-3}/\partial T_1$ is
$\displaystyle {{1\over 6}}
\cdot 2(2T_2) = \displaystyle {{1\over 3\mu ^2}}$ ,
and to  $\partial {\cal W}
^{(2)}_{-3}/\partial T_2$ is  $\displaystyle {{1\over 6}}
\cdot 4(T_1) = \displaystyle {
{2\over 3\mu }}$ . Also
$(2T_2){\cal W} ^{(2)}_{-3}$ contributes  $\displaystyle {{1\over 6}}
\cdot 2(T_1)(2T_2)^2 =
\displaystyle {{1\over 3\mu ^5}}$ ,
$(5T_5){\cal W} ^{(2)}_0 -  (5T_5)\cdot \displaystyle {{1\over 9}} =
\displaystyle {{1\over 9\mu ^5}}$ ,
$(2T_2){\cal W} ^{(2)}_0 -  (2T_2)\cdot \displaystyle {{1\over 9}} =
\displaystyle {{1\over 9\mu ^2}}$ ,  $(T_1){\cal W} ^{(2)}_{-3} -
\displaystyle {{1\over 6}}
\cdot 2(T_1)^2(2T_2) = \displaystyle {{1\over 3\mu ^4}}$ ,
$(4T_4){\cal W} ^{(2)}_0
-  (4T_4)\cdot \displaystyle {{1\over 9}} = \displaystyle {{1\over 9\mu ^4}}$
and  $(T_1){\cal W} ^{(2)}_0 -
(T_1)\cdot \displaystyle {{1\over 9}} = \displaystyle {{1\over 9\mu }}$ .
Putting this all together we obtain
the coefficient

$$
{1\over 27}\left\lbrace [4+1] + 9[{1\over 3} + {1\over 3} + {1\over 9} +
{1\over 9}] + 9[{2\over 3} + {1\over 3} + {1\over 9} +
{1\over 9}]\right\rbrace  +  O (\alpha ) = {28\over 27} +  O
(\alpha )
$$
in front of  $Z$  in (4.36), in accordance with (4.37). In order to reproduce
the term  $\displaystyle {{(K-1)(K+1)(2K+1)\over 24K}}
\alpha ^{K-2}Z  =  {7\over 9}
\alpha \ Z$  in (4.37) it is necessary to restore the shift (4.38) of
$T_4$-variable in (4.36). There are also contributions with  $T^2_4$ in
${\cal W} ^{(3)}$ and in $(3k-2)T_{3k-2}{\cal W} ^{(2)}_{3k+3n}$ , which are
responsible to occurrence of the $\alpha ^2$-term in (4.37).

This illustrative comparison of (4.36) and (4.37) is enough to recognize the
proper structure of (4.36). After it is found
out it is easy to ensure that all
the remaining contributions to (4.36) and (4.37)
also coincide. Of course, in
order to check the trace-structure of (4.36) our simple example of  $N = 1$  is
not enough --- the analogue of detailed consideration of sect.4.2 is required.
Such detailed derivation, as well as consideration of the case of  $K > 3$ , is
beyond the scope of this paper.

\section{Conclusion}

To conclude, we presented enough evidence that the Generalized Kontsevich Model
(1.2) interpolates between all the multimatrix models, while preserving the
property of integrability and the  ${\cal L} _{-1}$-constraint. This makes GKM
a very appealing candidate for the role of a theory, which could unify all
``stable" $(i.e$. with  $c \leq  1)$ bosonic string models. However, the study
of GKM was at most originated here. Let us list a set of problems, which seem
interesting for the future development.

1) Though integrable structure and  ${\cal L} _{-1}$- constraint in GKM have
been discussed more or less exhaustively, the situation with generic
${\cal W}$-constraints
and their implications remains less satisfactory. Of course, they
may be derived from  ${\cal L} _{-1}$-constraint, as suggested in
refs.\cite{FKN91},
but still it is desirable:

--- to complete the derivation of ${\cal W}$-constraints
from Ward identity (1.23) for
$N = \infty $, as suggested in sect.3, for  $K \geq  3$ ;

--- to prove the equivalence of these  ${\cal W}$-constraints
(which look like (1.26))
to conventional constraints (1.29);

--- to describe the subvariety $U_\infty$
in Grassmannian (at infinity of the Universal
module space), specified by the universal  ${\cal L} ^{\{{{\cal V}} \}}_{-1}$-
constraint;

--- to take double-scaling continuum limit of the properly reduced
$\tilde {\cal W}$-constraints in discrete multimatrix models (which look
like  $\tilde {\cal W} ^{(p+1)}_{q-p}[M_1] =
\tilde {\cal W}^{(q+1)}_{p-q}[M_2]$  for the 2-matrix case $[\tilde {\cal W}
])$ in order to derive the constraints (1.31) for
$\sqrt{\Gamma ^{\{K-1\}}_{ds}}$ (as it was done in ref.\cite{MMMM91} for
the 1-matrix case);

--- also the problems discussed in sect.3.4 should be added to this list.

2) We tried to argue, that particular  $(K-1)$-matrix model arises from GKM
for a particular choice of potential:  ${{\cal V}} (X) \equiv  {{\cal V}}_K (X)
=
const\cdot X^{K+1}$. Particular $(CFT + 2d$ gravity)-model is specified by
additional adjustment of  $M$-matrix in such a way, that all  $T_n= 0$ , $n
\neq  1$, $p+K$ (for $c = 1 - \displaystyle {{6(p - K)^2\over pK}})$.
These particular points
in the space of parameters  $\{{{\cal V}} (X)$, $M\}$  should be interpreted as
critical points of GKM. The relevant questions are:

--- what is the proper criterium, distinguishing critical points in terms of
GKM
itself;

--- are there any other critical points?

One can even hope, that the answer to the last question is positive and some
other critical points may be identified with non-bosonic string models. Let us
remind that the stability criterium like $c \leq  1$  (implying the absence of
tachionic instabilities, leading to desintegration of Riemann surfaces through
creation of holes), which in bosonic case implies the restriction  $d \leq  2$
for the space time dimension, in the case of  $N = 2$ superstrings turns into a
much more promising condition: $d \leq  4$ .

3) A separate class of problems is related to identification of
(double scaling limit)
$c=1$ models in the theory of GKM. This could lead to a more natural
formulation of GKM, in particular, help
to restore the symmetry between arguments
${{\cal V}}(X)$ and
$M$ of $Z^{\{{\cal V}\}}[M]$. Keeping in mind the topological
implications of $K=2$ model it is also natural to ask,

 --- what corresponds to
Penner model \cite{Pen87,DV91}.

4) The last group of problems concerns the meaning and generalizations of GKM.
First of all,
both formulations in terms of topological (like
\cite{Kon91,HZ86,Pen87,IZ90,DV91}) and conformal (in
the spirit of \cite{MMM91a,Ger91})
is required in order to understand explicitely
the connection to Liouville theory. Second,
the hope that just GKM (1.2) is enough to include  $N = 2$
superstrings may be too optimistic, and then one needs to look for more
universal models. Third, in any case one should search for a more invariant
(algebro-geometric) interpretation of (1.2): GKM is already much simpler
than original multi-matrix models (it is a 1-matrix model with a smooth
continuum limit), but still it is not simple enough to be appealing as a
fundamental theory.
The main aesthetic problem is the assymetry between $M$ and ${{\cal V}}(X)$
and,
perhaps, the very relevance of matrix integrals.
Fourth,
GKM does describe interpolation between sufficiently
many string models, but it is not a {\it dynamical} interpolation: one may
still need to look for an ``of-shell" version of GKM, involving a sort of
integration over  ${{\cal V}} (X)$.
This problem can be after all related to the
question about critical points. In particular, it may be interesting to find
out something similar to the ``off-shell" actions of
refs.\cite{GMMMMO91,JY90,GGPZ90}, which
reproduce Virasoro constraints in the ordinary 1-matrix models as dynamical
equations of motion, for the case of GKM. Possible links between GKM and
double-loop algebras (in the spirit of \cite{GLM90}) also deserve
investigation.

\section{Acknowledgments}

We are indebted for stimulating discussions to   A.Gerasimov, E.Gava,
M.Kontsevich, Yu.Makeenko, K.Narain, A.Orlov, S.Pakulyak, I.Zakharevich.


\begin{thebibliography}{10}

\bibitem{FKN91}
M.Fukuma, H.Kawai, and R.Nakayama, {\sl Int.J.Mod.Phys.}, {\bf A6} (1991) 1385;
{\it Infinite dimensional Grassmannian structure of two-dimensional quantum
gravity}, {\sl preprint~UT-572, KEK-TH-272} (1990); {\it Explicit solution for
$p$-$q$ duality in two-dimensional quantum gravity}, {\sl preprint~UT-582,
KEK-TH-289} (1991).


\bibitem{MMM91b}
A.Marshakov, A.Mironov, and A.Morozov, {\it On equivalence of topological and
  quantum 2d gravity}, {\sl preprint~HU-TFT-91-44, FIAN/TD/04-91, ITEP-M-4/91}
  (August, 1991).

\bibitem{Kaz89b}
V.A.Kazakov, {\sl Mod.Phys.Lett.}, {\bf A4} (1989) 2125.

\bibitem{BK90}
E.Br\'ezin and V.A.Kazakov, {\sl Phys.Lett.}, {\bf 236B} (1990) 144.

\bibitem{DS90}
M.Douglas and S.Shenker, {\sl Nucl.Phys.}, {\bf B335} (1990) 635.

\bibitem{GM90a}
D.Gross and A.A.Migdal, {\sl Phys.Rev.Lett.}, {\bf 64} (1990) 127.

\bibitem{CMM81}
S.Chadha, G.Mahoux, and M.Mehta, {\sl J.of Phys.}, {\bf A14} (1981) 579.

\bibitem{BPZ84}
A.A.Belavin, A.M.Polyakov, and A.B.Zamolodchikov, {\sl Nucl.Phys.},
{\bf B241} (1984)
  333.

\bibitem{Dou90}
M.Douglas, {\sl Phys.Lett.}, {\bf 238B} (1990) 176.

\bibitem{Sei90}
N.Seiberg, {\it Notes on quantum Liouville theory and quantum gravity}, {\sl
  preprint~RU-90-29} (1990).

\bibitem{KMMMZ91a}
S.Kharchev {\it et al.} {\it Unification
  of all string models with $c<1$}, {\sl preprint~FIAN/TD/9-91, ITEP-M-8/91}
  (October, 1991).

\bibitem{GN91}
D.J.Gross and M.J.Newman, {\sl Phys.Lett.}, {\bf 266B} (1991) 291-297.

\bibitem{MS91}
Yu.Makeenko and G.Semenoff, {\it Properties of Hermitean matrix model in an
  external field}, {\sl preprint~ITEP/UBC} (July, 1991).

\bibitem{MMM91c}
A.Marshakov, A.Mironov, and A.Morozov, {\it From Virasoro constraints in
  Kontsevich's model to ${\cal W}$-constraints in 2-matrix models}, {\sl
  preprint~FIAN/TD/05-91, ITEP-M-7/91} (August, 1991).

\bibitem{Kon91}
M.Kontsevich, {\sl Funk.Anal.\&Prilozh.}, {\bf 25} (1991) 50 (in Russian).

\bibitem{MMMM91}
Yu.Makeenko {\it et al.} {\sl Nucl.Phys.}, {\bf
  B356} (1991) 574.

\bibitem{GMMMMO91}
A.Gerasimov {\it et al.} {\sl
  Mod.Phys.Lett.}, {\bf A6} (1991) 3079.

\bibitem{Wit90}
E.Witten, {\sl Nucl.Phys.}, {\bf B340} (1990) 281.

\bibitem{Wit91}
E.Witten, {\it On the Kontsevich model and other models of two-dimensional
  gravity}, {\sl preprint~IASSNS-HEP-91/24} (June, 1991).

\bibitem{DVV91a}
R.Dijkgraaf, H.Verlinde, and E.Verlinde, {\sl Nucl.Phys.}, {\bf B348} (1991)
  435.

\bibitem{Orl88}
A.Yu.Orlov, In: {\it Proc.Kiev Int.Workshop}, page~116, World
  Scientific, Singapore, 1988.

\bibitem{Fut}
{\it W-constraints in discrete multi-matrix models}, {\sl preprint~FIAN-ITEP}
(in
  preparation).

\bibitem{KS91}
V.Kac and A.S.Schwarz, {\sl Phys.Lett.}, {\bf B257} (1991) 329.

\bibitem{IZ80}
C.Itzyzson and J.-B.Zuber, {\sl J.Math.Phys.}, {\bf 21} (1980) 411.

\bibitem{Meh81}
M.L.Mehta, {\sl Commun.Math.Phys.}, {\bf 79} (1981) 327.

\bibitem{Miw82}
T.Miwa, {\sl Proceedings of the Japan Academy}, {\bf 58} (1982) 9.

\bibitem{KMMOZ91}
S.Kharchev {\it et al.} {\it Matrix models
  among integrable theories: forced hierarchies and operator formalism}, {\sl
  preprint~FIAN/TD/1-91} (February, 1991).

\bibitem{GMMMO91}
A.Gerasimov {\it et al.} {\sl Nucl.Phys.},
  {\bf B357} (1991) 565.

\bibitem{Sat81}
N.Sato, {\sl RIMS Kokyuroku}, {\bf 439} (1981) 30.

\bibitem{SW85}
G.Segal and G.Wilson, {\sl Publ.I.H.E.S.}, {\bf 61} (1985) 1.

\bibitem{Zab89}
A.Zabrodin, {\sl Teor.Mat.Phys.}, {\bf 78} (1989) 234.

\bibitem{DrS84}
V.Drinfeld and V.Sokolov, {\sl J.Sov.Math.}, {\bf 30} (1984) 1975.

\bibitem{DJKM83}
E.Date {\it et al.} In: {\it Proc.RIMS
symp.Nonlinear integrable systems ---
  classical theory and quantum theory}, page~39, Kyoto, 1983.

\bibitem{GM90b}
D.Gross and A.A.Migdal, {\sl Nucl.Phys.}, {\bf B340} (1990) 333.

\bibitem{OS86}
A.Orlov and E.Schulman, {\sl Lett.Math.Phys.}, {\bf 12} (1986) 171.

\bibitem{GO89}
P.Grinevich and A.Orlov, {\it Flag spaces in KP theory and Virasoro action on
  $\det \tilde \partial _j$ and Segal-Wilson $\tau$-function},
  {\sl preprint~Cornell Univ.}
  (September, 1989).

\bibitem{Sem89}
A.Semikhatov, {\sl Int.J.Mod.Phys.}, {\bf A4} (1989) 467.

\bibitem{Pen87}
R.Penner, {\sl J.Diff.Geom.}, {\bf 27} (1987) 35.

\bibitem{DV91}
J.Distler and C.Vafa, {\sl Mod.Phys.Lett.}, {\bf A6} (1991) 259.

\bibitem{HZ86}
J.Harer and D.Zagier, {\sl Invent.Math.}, {\bf 85} (1986) 457.

\bibitem{IZ90}
C.Itzykson and J.-B.Zuber, {\it Matrix integration and combinatorics of modular
  groups}, {\sl preprint~SPhT/90-004} (1990).

\bibitem{MMM91a}
A.Marshakov, A.Mironov, and A.Morozov, {\sl Phys.Lett.}, {\bf 265B} (1991) 99.

\bibitem{Ger91}
A.Gerasimov, {\sl preprint~ITEP} (1991).

\bibitem{JY90}
A.Jevicki and T.Yoneya, {\sl Mod.Phys.Lett.}, {\bf A5} (1990) 1615.

\bibitem{GGPZ90}
P.Ginsparg {\it al.} {\sl Nucl.Phys.},
{\bf B342}
  (1990) 539.

\bibitem{GLM90}
A.Gerasimov, D.Lebedev, and A.Morozov, {\it On possible implications of
  integrable systems to string theory}, {\sl preprint~ITEP-4-90} (1990).

\end{thebibliography}
\end{document}